\documentclass[]{aa}

\usepackage{graphicx}
\usepackage{color}

\begin{document}

\title{The far-infrared - radio correlation in star-forming dwarf galaxies}

\titlerunning{The far-infrared - radio correlation in dwarf galaxies}
\authorrunning{Schleicher \& Beck}

\author
  {Dominik R.G. Schleicher
  \inst{1}
  \and
  Rainer Beck
  \inst{2}
  }

\institute{Departamento de Astronom\'ia, Facultad Ciencias F\'isicas y Matem\'aticas, Universidad de Concepci\'on, Av. Esteban Iturra s/n Barrio Universitario, Casilla 160-C, Concepci\'on, Chile; 
\email{dschleicher@astro-udec.cl}
\and
Max-Planck-Institut f\"ur Radioastronomie, Auf dem H\"ugel 69, 53121 Bonn, Germany\\
\email{rbeck@mpifr-bonn.mpg.de}
}

\date{\today}

\abstract{
The far-infrared - radio correlation connects star formation and magnetic fields in galaxies, and has been confirmed over a large range of far-infrared / radio luminosities, both in the local Universe and even at redshifts of $z\sim2$. Recent investigations indicate that it may even hold in the regime of local dwarf galaxies, and we therefore explore here the expected behavior in the regime of star formation {surface densities} below $0.1$~M$_\odot$~kpc$^{-2}$~yr$^{-1}$. We derive two conditions that can be particularly relevant for inducing a change in the expected correlation: a critical star formation surface density to maintain the correlation between star formation rate and the magnetic field, and a critical star formation surface density below which cosmic ray diffusion losses dominate over their injection via supernova explosions. For rotation periods shorter than
$1.5\times10^7 (H/\mathrm{kpc})^2$~yrs, with $H$ the scale height of the disk, the first correlation will break down before diffusion losses are relevant, as higher star formation rates are required to maintain the correlation between star formation rate and magnetic field strength. For high star formation surface densities $\Sigma_{\rm SFR}$, we derive a characteristic scaling of the non-thermal radio to the far-infrared / infrared emission with $\Sigma_{\rm SFR}^{1/3}$, corresponding to a scaling of the non-thermal radio luminosity $L_s$ with the infrared luminosity $L_{th}$ as $L_{th}^{4/3}$. The latter is expected to change when the above processes are no longer steadily maintained. In the regime of long rotation periods, we  expect a transition towards a steeper scaling with $\Sigma_{\rm SFR}^{2/3}$, implying $L_s\propto L_{th}^{5/3}$, while the regime of fast rotation is expected to show a considerably enhanced scatter, as a well-defined relation between star formation and magnetic field strength is not maintained.  The scaling relations above explain the increasing thermal fraction of the radio emission observed within local dwarfs, and can be tested with future observations by LOFAR as well as the SKA and its precursor radio telescopes.
}

\maketitle

\section{Introduction}
Over the last years, magnetic fields have been detected in a significant number of local dwarf galaxies. This includes prominent examples such as the Large Magellanic Cloud \citep[LMC, ][]{Gaensler05}, the Small Magellanic Cloud \citep[SMC, ][]{Mao08}, and many additional examples such as NGC~4449 \citep{Chyzy00}, NGC~ 1569 \citep{Kepley10}, NGC~6822 \citep{Chyzy03}, IC~10 \citep{Chyzy03, Heesen11} and NGC~ 4214 \citep{Kepley11}. \citet{Chyzy11} pursued a dedicated investigation of radio emission and magnetic fields in an unbiased sample of $12$ Local Group (LG) irregular and dwarf irregular galaxies yielding both detections and upper limits, while \citet{Roychowdhury12} employed the stacking technique to improve the sensitivity in the radio and to probe average properties of the radio emission for the faintest end of dwarf galaxies. A central result of both studies is that the magnetic fields in dwarf galaxies are about three times weaker than in normal spirals, with a typical field strength of $<4.2\pm1.8$~$\mu$G as given by \citet{Chyzy11}. Both the detections and  upper limits are consistent with the assumption that local dwarf galaxies lie on the far-infrared - radio correlation, with a typical scaling of the magnetic field strength $B$ with the star formation surface density $\Sigma_{\rm SFR}^{1/3}$.

The correlation between the far-infrared and radio fluxes was originally observed by \citet{Kruit73b, Kruit73c, Kruit73a}. Subsequent investigations have been pursued by \citet{deJong85} and \citet{Helou85}, while the interpretation  in terms of calorimeter models was pursued by \citet{Volk89}. \citet{Niklas97b} proposed a detailed scenario in which the far-infrared - radio correlation emerges due to a relation between the magnetic field strength, the gas surface density and the star formation rate. In particular, it is well-known that the gas surface density is strongly correlated to star formation activity, as reflected in the Kennicutt-Schmidt relation \citep{Schmidt59, Kennicutt98, Kennicutt08, Walter08, Bigiel11, Kennicutt12}. Massive stars emit UV radiation absorbed by dust grains, and re-emitted in the infrared and far-infrared. In addition, the supernova explosions of massive stars inject both cosmic rays and turbulence into the interstellar medium. Such turbulence efficiently amplifies magnetic field via the small-scale dynamo \citep{Kazantsev68, Subramanian99, Scheko02, Schober12b, Schleicher13, FederrathPRL, Grete15}, and as a result, the feedback from star formation provides the relevant ingredients to drive the radio emission \citep[see e.g.][]{Groves03, Schleicher13b}. 

The potential validity of the far-infrared - radio correlation even in dwarf galaxies was already suggested by \citet{Bell03}, as both the dust content in the dwarfs and the efficiency of non-thermal radio emission may decrease in a similar amount towards lower star formation rates. At that time, the correlation had been observationally established only for nearby spiral galaxies, including a sample of 1809 galaxies probed by \citet{Yun01}, for which the correlation has been confirmed over 5 orders of magnitude in luminosity. More recent work by \citet{Lacki10} has shown that in fact both the escape of UV photons and cosmic rays from the galaxy may be correlated to the characteristic surface densities, and contributes to our overall understanding of the observations. It is  worth noting that these correlations do not only hold on a global scale, but have further been confirmed within the galaxies via dedicated investigations \citep{Dumas11, Taba13, Heesen14}.

Recent studies by \citet{Murphy09}, \citet{Ivison10a}, \citet{Jarvis10}, \citet{Sargent10} and \citet{Casey12} in fact provide evidence that the far-infrared - radio correlation holds at least until redshifts of $z\sim2$, and it also holds in the context of galaxy mergers \citep{Drzazga11}.  {In addition, work by \citet{Miettinen15} shows that the radio emitting region is more extended than the infrared emitting region at least in some cases. The latter is potentially consistent with Taffy-like systems \citep{Condon02}, mergers \citep{Murphy13} or systems undergoing tidal interactions \citep{Donevski15}, though we suggest that tidal tails as in the Antennae galaxies \citep{Chyzy04} may be the more frequent scenario. In the context of mergers, previous studies, e.g. \citet{Drzazga11} have preferentially considered the impact of magnetic fields, while \citet{Lisenfeld10} pointed out the potential importance of particle acceleration. The latter requires rather high Mach numbers, which may not be available in the interstellar medium \citep{Guo14a}, or the Firehose instability, if the acceleration only concerns the electrons \citep{Guo14b}. In that case, a high plasma beta would be required, while current estimates indicate that it may be rather small \citep{Beck15}.}

The radio-infrared correlation holds for thermal and non-thermal (synchrotron) radio emission, though with different slopes. Thermal radio emission may dominate in dwarf galaxies \citep{Roychowdhury12} and hence mask the relation between non-thermal and infrared emission. The interpretation requires a careful separation of both emission components, e.g. with the help of spectral index data, which is non-trivial and often not possible. Alternatively, the thermal radio emission is assumed to be linearly proportional to the infrared emission, although not all infrared emission is directly related to UV radiation from young stars. Only H$\alpha$ emission can be safely assumed to be proportional to radio thermal emission, if extinction is properly corrected.

On theoretical grounds, it is  expected that magnetic fields can be efficiently ampified even in small systems and at high redshift, due to the efficient amplification by turbulence \citep{Arshakian09, Wang09, Schleicher10c, Souza10, Latif13, Schober13}, and it is thus conceivable that the far-infrared - radio correlation will be in place early on. As pointed out by \citet{Murphy09}, a potential difficulty is however the increasing strength of the cosmic microwave background at high redshift, enhancing the inverse Compton emission and providing an additional loss mechanism for the cosmic ray electrons. It is thus conceivable that the latter may lead to a modification or a breakdown of the correlation at very high redshift due to differences in the energy loss mechanisms of cosmic rays \citep{Lacki10b, Schleicher13b, Schober15}.

In this paper, we explore whether a correlation between the far-infrared and radio emission can still be expected in dwarf galaxies, in particular in the limit of star formation {surface densities} below $0.1$~M$_\odot$~kpc$^{-2}$~yr$^{-1}$. In this respect, one needs to examine a couple of relevant issues. A first one concerns the behavior of star formation itself. In the sample explored by \citet{Chyzy11}, the Kennicutt-Schmidt relation appeared to hold even in the dwarf galaxy regime, while \citet{Roychowdhury09} reported potential deviations towards lower star formation rates. Through dedicated spatially resolved investigations of star formation and the HI-dominated gas both in nearby spirals and dwarf irregular galaxies using the THINGS \citep{Walter08} and FIGGS \citep{Begum08} survey, \citet{Roychowdhury15} have pursued a detailed comparison the the Kennicutt-Schmidt relation in both regimes, finding in particular that there is no dependence on the metallicity of the gas. It is therefore conceivable that this relation will hold in the dwarf galaxy regime, even though the scatter may potentially increase towards lower gas masses.

An important difference compared to spiral galaxies is that the rotation in local dwarfs may be substantially reduced, implying slow or more chaotic rotation with a low differential rotation \citep{Chyzy03}. In the VLA-ANGST survey of 35 nearby dwarf galaxies, the lowest detected rotation velocities have been of the order $20$~km/s.  Through the Westerbork HI Survey of Spiral and Irregular Galaxies (WHISP), rotation curves have been measured for a sample of 62 galaxies, finding typical rotational velocities of $20-80$~km/s on scales of a few disk scale lengths \citep{Swaters09}. In general, a strong variation is found from dwarf to dwarf, and in some cases like NGC4449 \citep{Theis01} or IC~10 \citep{Ashley14}, the rotation curves have been explained by tidal interactions of disks with other dwarfs.  The implications of the large diversity of dwarf galaxy rotation curves has therefore been discussed also in recent studies \citep{Oman15}.

In the radio regime, additional differences have been found compared to the typical conditions in local spirals, whose main properties were already described by \citet{Condon92}. In particular, while the fraction of thermal radio emission corresponds to about $8\%$ for local spirals \citep{Murphy06}, a non-thermal fraction of $\sim50\%$ has been reported recently by \citet{Roychowdhury12} in the dwarf galaxy regime. We will in fact show in this paper that such a behavior arises rather naturally due to the non-linear nature of the far-infrared - radio correlation. 

The structure of this paper is as follows. In section~\ref{model}, we outline our overall modelling framework, which is employed to derive characteristic timescales both for dynamical processes within the dwarfs as well as the timescales for thermal and radio emission. These are employed to derive critical star formation surface densities, which are required for these processes to be maintained in a steady fashion. In section~\ref{slope}, these results are employed to distinguish between four characteristic regimes of radio emission in the dwarf galaxies, and we discuss in particular the expected slope and the breakdown of the correlation at very low star formation rates. A discussion with our main conclusions is presented in section~\ref{discussions}.

\section{Model framework}\label{model}
In the following, we will outline our main model framework, starting with the basic assumptions (subsection~\ref{basic}) and including a more detailed framework for the formation of galactic winds (subsection~\ref{windmodel}). A discussion of relevant timescales for the dynamics, the thermal and radio emission is given in subsection~\ref{timescales} along with a derivation of critical star formation rates, which are presented and compared in subsection~\ref{criticalsfr}.

\subsection{Basic assumptions}\label{basic}
We assume that star formation in  dwarf galaxies occurs in a disk of neutral HI gas with surface density $\Sigma$, radius $R_{\rm disk}$ and height $H$. We further assume that the Kennicutt- Schmidt relation provides a valid description of the star formation law, so that the star formation surface density is given as\begin{equation}
\Sigma_{\rm SFR}=C\Sigma^N,\label{Kennicutt}
\end{equation}
where $C$ is a normalization constant and $N\sim1.5$ the slope of the relation. Considering the results by \citet{Chyzy11}, we adopt here \begin{equation}
C\sim\frac{2\times10^{-2}\ M_\odot\ \mathrm{yr}^{-1}\ \mathrm{kpc}^{-2}}{\left( 10^8\ M_\odot\ \mathrm{kpc}^{-2} \right)^N}
\end{equation}
for the normalization. With these quantities, the amount of gas available for star formation is $M_{gas}\sim R_{\rm disk}^2\pi\Sigma$, and the total star formation rate within the galaxy is given as\begin{equation}
\dot{M}_{\rm SFR}=R_{\rm disk}^2\pi \Sigma_{\rm SFR}=R_{\rm disk}^2\pi C\Sigma^N.
\end{equation}
The observed relation between non-thermal radio and infrared emission further suggest a relation {between the magnetic field strength $B$ and the star formation surface density $\Sigma_{\rm SFR}$ of the form}\begin{equation}
B=C_B \Sigma_{\rm SFR}^{1/3},\label{BSFR}
\end{equation}
where we determine the normalization constant $C_B$ from the investigation of \citet{Chyzy11} as\begin{equation}
C_B\sim\frac{8\mu G}{\left( 0.1\ M_\odot\ \mathrm{yr}^{-1}\ \mathrm{kpc}^{-2} \right)^{1/3}}.
\end{equation}

The height of the disk of the warm interstellar medium (ISM) results from the balance between the gravitational force, the turbulent pressure, thermal pressure, magnetic pressure and cosmic rays. The disk height $H$ is then given as\begin{equation}
H = \frac{v_{th}+v_t(1+2\epsilon_{\rm B}^{1/2})}{\Omega_K},
\end{equation}
where $v_{th}\sim10$~km/s corresponds to the thermal velocity of the warm ISM, $v_t$ is the turbulent velocity and $\Omega_K$ the angular velocity required to balance the gravitational force. In general, we note that the disk height will be different depending on the ISM component that is considered, as the cold gas may have different turbulent and lower thermal velocities. We assume here equipartition between the magnetic energy density and cosmic rays, with the magnetic energy density given as a fraction $\epsilon_{\rm B}$ of the turbulent energy density. Indeed, observations have shown that the energy density of turbulence and the magnetic field are essentially comparable, implying $\epsilon_{\rm B}\sim1$ \citep{Beck07, Beck15}. The latter indicates that turbulence may play a central role in amplifying the magnetic field. 

Also in this paper, we will in the following assume approximate equipartition between magnetic and turbulent energy in the warm gas component, i.e. the component which predominantly gives rise to the observed radio emission. {We particularly note that we focus here on the disordered component of the magnetic field, which yields the dominant contribution to the radio flux. For comparison, studies by \citet{Drzazga16} have shown that the ordered component in the outer parts of the galaxy is also affected by tidal interactions.} The density of the warm neutral gas can be estimated as $\Sigma/2H$. The typical density ratio between the ionized and neutral warm gas component is denoted as $C_{\rm HII/HI}$ in the following, and we adopt a typical value of $C_{\rm HII/HI}\sim2\%$ following \citet{Tielens05}. The equipartition between turbulent and magnetic energy in the warm ionized gas can then be expressed as
\begin{equation}
\frac{B^2}{8\pi}=\epsilon_{\rm B}C_{\rm HII/HI}\frac{1}{2}\frac{\Sigma}{2H}v_t^2.\label{equi}
\end{equation}
With the {order-of-magnitude} scaling relation $H\sim v_t/\Omega_K$, the above expression can be simplified as
\begin{equation}
\frac{B^2}{8\pi}=\frac{1}{4}\epsilon_{\rm B}C_{\rm HII/HI}\Sigma v_t \Omega_K.\label{equi}
\end{equation}
We can therefore solve for the turbulent velocity $v_t$, and insert  Eq.~(\ref{BSFR}) as well as the Kennicutt-Schmidt relation (Eq.~\ref{Kennicutt}), yielding\begin{eqnarray}
v_t&=&\frac{B^2}{2\pi\epsilon_{\rm B}C_{\rm HII/HI}\Sigma\Omega_K}=\frac{C_B^2\Sigma_{\rm SFR}^{2/3}}{2\pi\epsilon_{\rm B}C_{\rm HII/HI}\Sigma\Omega_K}\nonumber\\
&=&\frac{C_B^2 C^{2/3}\Sigma^{2N/3-1}}{2\pi\epsilon_{\rm B}C_{\rm HII/HI}\Omega_K}.\label{vt}
\end{eqnarray}
Evaluating the disk height and turbulent velocity in this way, additional properties of the galaxy follow from the more detailed ISM model described in the next subsection. A central parameter is then the amount of rotation $\Omega_K$ of the dwarf galaxy.

\subsection{Wind model}\label{windmodel}
\begin{figure*}[htbp]
\begin{center}
\includegraphics[scale=1]{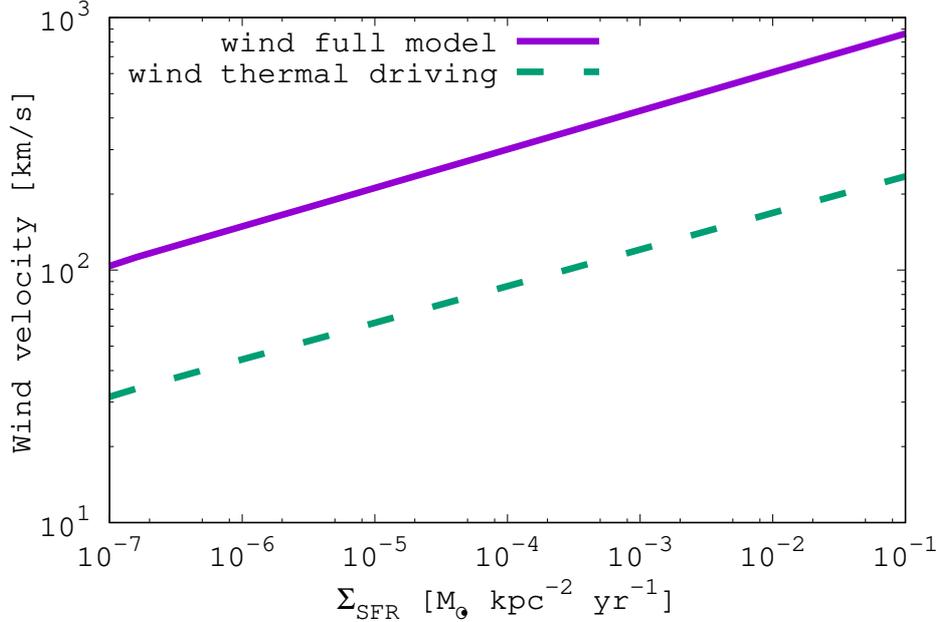}
\caption{Wind velocity as a function of the star formation surface density. Shown are both the velocities for a thermally-driven wind (Eq.~\ref{windtherm}) and our full model including magnetic and cosmic-ray pressure (Eq.~\ref{wind}). As real dwarf galaxies show a relevant scatter in their typical field strength, we expect realistic cases to lie inbetween.}
\label{figwind}
\end{center}
\end{figure*}

The wind model adopted here is an extension of the framework developed by \citet{Shu05}. They employed the multi-phase ISM model for supernova evolution by \citet{McKee77} and \citet{Efstathiou00} to construct a model for galactic winds in the context of starburst galaxies. A central assumption in their model is thus that the galactic porosity, i.e. the volume filling factor of the hot gas, which is formally defined via\begin{equation}
P=\frac{f_d V_{\rm hot}}{V_{\rm SFR}},
\end{equation}
is generally of order $1$, with $V_{\rm hot}$ the volume of the hot gas, $V_{\rm SFR}$ the total volume of the star-forming system and $f_d\sim 2H/R$ a correction factor for galactic disks. To generalize this point, we therefore adopt the expression derived by \citet{Clarke02} to parametrize the dependence of porosity on galaxy properties, which is given as\begin{equation}
P=\frac{7 f_d (\dot{M}_{\rm SFR}/(M_\odot/yr))}{(M_{\rm ISM}/10^{10}~M_\odot)(v_{th}/(10\ km/s))^2}.
\end{equation}
In the above, $M_{\rm ISM}$ denotes the mass of the ISM, which we estimate via $M_{ISM}=R_{\rm disk}^2\pi \Sigma${, and $\dot{M}_{\rm SFR}$ denotes the star formation rate in the galaxy}. The stellar mass here is not considered, as we only divide by the thermal energy of the warm ISM. The warm gas produced due to star formation is compared with the thermal energy of the warm ISM, adopting a thermal velocity of $v_{th}\sim10$~km/s. Unless in the regime of extremely low star formation rates, we generally find $P\sim1$, consistent with previous results by \citet{Clarke02}.

 {We note that the expressions adopted here assume a Salpeter IMF \citep{Salpeter55}, consistent with the assumption that about $5\%$ of the stellar mass is in massive stars with more than $8$~M$_\odot$. The precise form of the IMF in dwarf galaxies is however a matter of ongoing debate. For instance, the results by \citet{Geha13} for ultra-faint dwarf galaxies indicate that the IMF is more shallow than Salpeter in the mass range between $0.52$~M$_\odot$ and $0.77$~M$_\odot$. On the other hand, a comparison of H$\alpha$ and UV data suggests a deficient of high-mass stars below star formation rates of $0.003$~M$_\odot$~yr$^{-1}$ \citep{Lee09}, in line with expectations of \citet{Weidner05}. For a generalization of the framework adopted here, all star formation rates or surface densities in the following may be considered to be multiplied by a factor $\epsilon_{hm}/0.05$. We note here in particular that a lower value of $\epsilon_{hm}$ may imply that some of the relations discussed here could break down even earlier, or, in case of a gradual evolution of this parameter, a steepening of the far-infrared - radio relation would occur. In the absence of a more detailed knowledge, we will in the following however assume that the parameter is constant, and explore the corresponding results.}

\begin{figure*}[htbp]
\begin{center}
\includegraphics[scale=1]{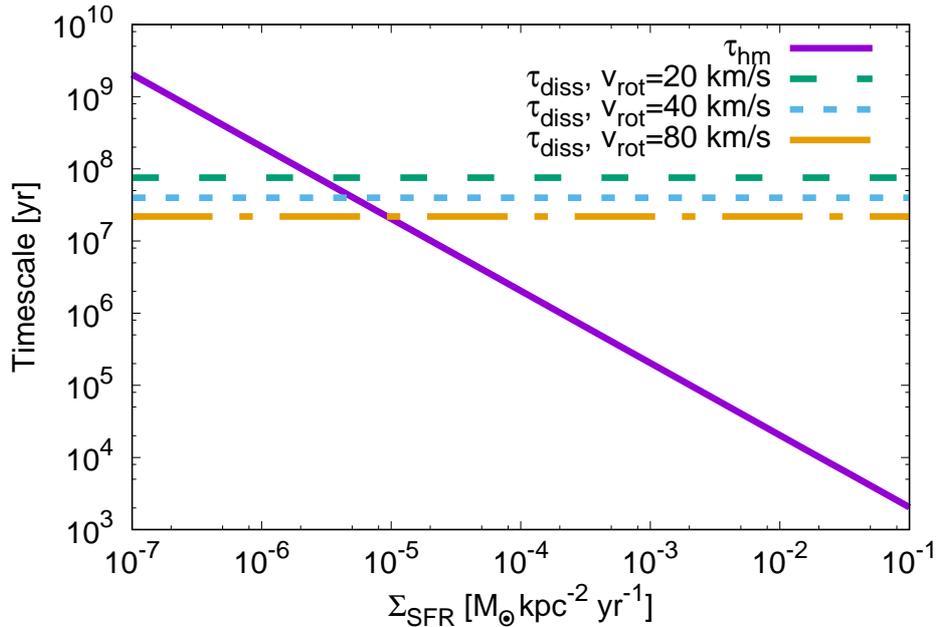}
\caption{Turbulence injection timescale (or timescale of massive star formation) {compared to} turbulence dissipation timescale in dwarf galaxies with rotational velocities of $20$~km/s, $40$~km/s and $80$~km/s. The timescales are shown as a function of star formation surface density. The critical star formation surface density is found to be $\sim10^{-5}$~M$_\odot$~kpc$^{-2}$~yr$^{-1}$, with only a weak dependence of the rotational velocity. The size of the star-forming region is assumed to be $0.5$~kpc.}
\label{sfr_diss}
\end{center}
\end{figure*}

From the star formation surface density $\Sigma_{\rm SFR}$ in the galaxy, we can estimate the star formation rate per unit volume as $\dot{\rho}_{\rm SFR}=\Sigma_{\rm SFR}/2H$. For the model of \citet{Shu05}, we first need to calculate the supernova rate $S_{-13}$ in units of $10^{-13}$~pc$^{-3}$~yr$^{-1}$. The latter is given as\begin{equation}
S_{-13}=10^{13}\frac{\dot{\rho}_{\rm SFR}}{M_{ps}}=\frac{10^{13}\times\Sigma_{\rm SFR}}{2HM_{ps}},
\end{equation}
where $M_{ps}$ denotes the required stellar mass to form at least one star massive enough to explode as a supernova. The latter is parametrized via $M_{ps}=8$~M$_\odot/\epsilon_{hm}$, with $\epsilon_{hm}\sim0.05$ the mass fraction of massive stars \citep{Kroupa02, Chabrier03}. From this expression, one can calculate the temperature of the hot gas in cavities as \citep{Efstathiou00, Shu05}:\begin{equation}
T_h=6.6\times10^5\left( \frac{ S_{-13}E_{51}f_\Sigma}{\gamma}\right)^{0.29} K,\label{Th}
\end{equation}
with $E_{51}=1$ the typical energy of supernova explosions in units of $10^{51}$~erg and $\gamma=2.5$ the ratio of blast wave velocity to the isothermal sound speed of the hot phase in the case of strong shocks. The quantity $f_\Sigma$ parametrizes the dependence on various properties of the ISM, including conductivity and minimum mass of the clouds. For most of them, we keep the standard parameters employed by \citet{Shu05}, and consider in the following only the impact of the varying cold gas fraction $f_c=e^{-P}$. The latter yields\begin{equation}
f_\Sigma = 21.5\left( \frac{f_c}{e^{-1}} \right)^{-1}.
\end{equation}
We note here that the final wind velocity will scale as $f_\Sigma^{0.145}$, and is therefore highly insensitive to the ISM parameters incorporated into $f_\Sigma$. In a fully ionized gas, the isothermal sound speed is now given as\begin{equation}
C_i=\sqrt{k_B T_h/\mu m_p}=37 \left( \frac{T_h}{10^5\ K} \right)^{0.5}\ \mathrm{km/s},\label{Ci}
\end{equation}
with $k_B$ the Boltzmann constant, $m_p$ the proton mass and $\mu$ the mean molecular weight. In the presence of a pure thermal driving, the wind velocity then follows as\begin{equation}
v_{\rm wind, therm}=\Gamma_W C_i P^{-1/7},\label{windtherm}
\end{equation}
where the factor $\Gamma_W\sim\sqrt{2.5}$ allows for some radiative cooling of the energy injected by the supernova, and the factor $P^{-1/7}$ provides a correction in the limit of low porosity parameters developed by \citet{Shu05}.

Here, we further aim to account for the role of magnetic and cosmic ray pressure in driving such winds. For this purpose, we note that Eq.~(\ref{windtherm}) has been derived from the conservation of the specific enthalpy in the hot gas. We recall that the thermal enthalpy of the hot gas is given as $H_T=\frac{5}{3}E_T$, with $E_T$ its thermal energy. The magnetic enthalpy is given as $H_B=2E_B$, with $E_B$ the magnetic energy, and the enthalpy of cosmic rays follows as $H_{CR}=\frac{4}{3}E_{CR}$, with $E_{CR}$ the cosmic ray energy. We assume here approximate equipartition between the energy in magnetic fields and cosmic rays, and our expression for the total enthalpy is then given as \begin{equation}
H_{\rm tot}=\frac{5}{3}E_T\left(1+2\frac{E_B}{E_T}  \right).
\end{equation}
The ratio $E_B/E_T$ is  evaluated using\begin{equation}
\frac{E_B}{E_T}=\frac{B^2/8\pi}{1.5 n_h k_B T_h},
\end{equation}
where $n_h$ is the number density of the hot gas, which we evaluate as \citep{Efstathiou00}\begin{equation}
n_h=4.3\times10^{-3}S_{-13}^{0.36}\gamma^{-0.36}f_\Sigma^{-0.393}.\label{nh}
\end{equation}
Considering thus the conservation of thermal, magnetic and cosmic ray enthalpy, the resulting wind velocity is given as\begin{equation}
v_{\rm wind, therm}=\Gamma_W C_i P^{-1/7}\sqrt{1+2\frac{E_B}{E_T}}.\label{wind}
\end{equation}
The resulting wind velocities both for the full model and the thermally-driven wind are illustrated in Fig.~\ref{figwind}. {To understand the resulting behavior as a function of the star formation surface densitiy, we first note that the porosity $P$ is close to $1$ for most of the regime considered here, and we also assume a constant ratio $E_B/E_T$. The main variable through which the star formation surface density enters in the calculation is thus the isothermal sound speed $C_i$ given in Eq.~(\ref{Ci}), which depends on the temperature of the hot gas in ISM cavities given in Eq.~(\ref{Th}). The latter scales with $S_{-13}^{0.29}$, with $S_{-13}\propto \Sigma_{\rm SFR}/H$. For a typical case with Kennicutt-Schmidt index $N=1$, we have $H\sim v_t/\Omega_K\propto \Sigma_{\rm SFR}^0/\Omega_K$ from Eq.~(\ref{vt}), where $\Omega_K$ is independent of the star formation rate. Thus, $T_h\propto \Sigma_{\rm SFR}^{0.29}$ and $v_{\rm wind}\propto C_i\propto C_i\propto T_h^{0.5}\propto \Sigma_{\rm SFR}^{0.145}$. The latter implies that the wind velocity changes over one order of magnitude when the star formation surface density changes by 7 orders of magnitude, consistent with the results in Fig.~\ref{figwind}.  }

\begin{figure*}[htbp]
\begin{center}
\includegraphics[scale=1]{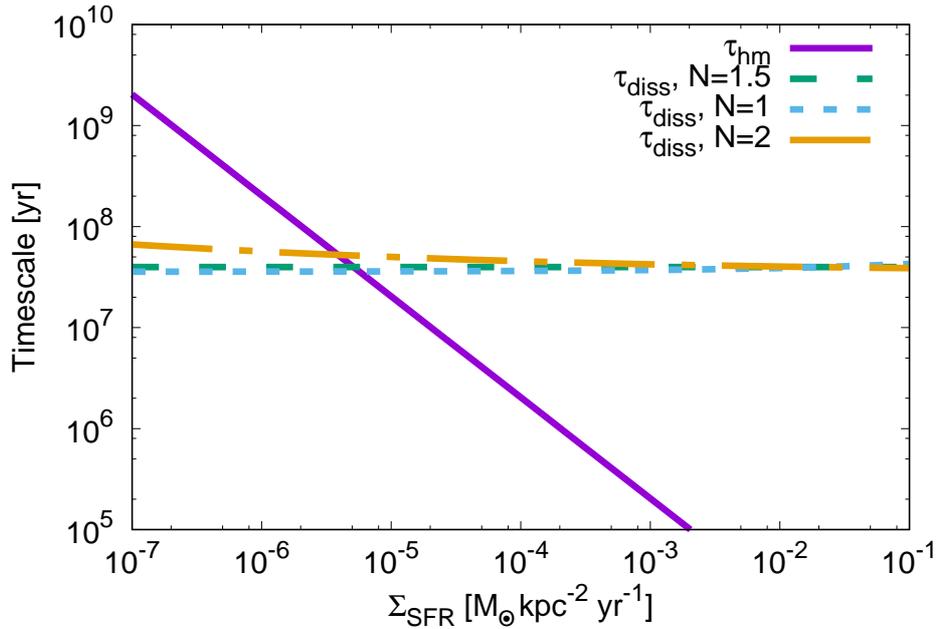}
\caption{Turbulence injection timescale (or timescale of massive star formation) {compared to} turbulence dissipation timescale in dwarf galaxies for different Kennicutt-Schmidt relations {as a function of the star formation surface density}. We adopt here a reference case with $N=1.5$ as well as two additional cases with $N=1$ and $N=2$. The size of the star forming region is assumed to be $0.5$~kpc, the rotational velocity $40$~km/s. We find that the critical star formation surface density has no strong dependence on the precise form of the Kennicutt-Schmidt relation.}
\label{sfr_var}
\end{center}
\end{figure*}

\subsection{Characteristic timescales}\label{timescales}
In this subsection, we  evaluate characteristic timescales of the dwarf galaxies and their ISM to assess whether the far-infrared - radio relation can be maintained for low star formation rates. In the following, we  distinguish between dynamical timescales, the timescale for the thermal emission and the timescales for radio emission and cosmic ray losses.

\subsubsection{Dynamical timescales}

Assuming that the disk height $H$  corresponds to the size of the largest turbulent eddies, the timescale for the turbulent energy dissipation
 is given as\begin{equation}
\tau_{\rm diss}\sim\frac{H}{v_t}\sim \Omega_K^{-1}.
\end{equation}
The balance between gravitational force and turbulent pressure within the vertical direction therefore implies that the turbulence dissipation time is comparable to the rotation period of the dwarf galaxy $\Omega_K^{-1}$, assuming that the vertical support of the disk is dominated by the turbulent energy, i.e. $H\sim \Omega_K/v_t$. We do not explicitly account here for the potential effects of differential rotation, which may change the rotation period as a function of radius, as we are predominantly interested in an order-of-magnitude estimate. {For $\tau_{\rm diss}$, we expect overall a very weak dependence on the star formation surface density, which in fact vanishes under the assumption that $H\sim v_t/\Omega_K$. Such a very weak dependence is indeed visible in Fig.~\ref{sfr_diss}.}

\begin{figure*}[htbp]
\begin{center}
\includegraphics[scale=1]{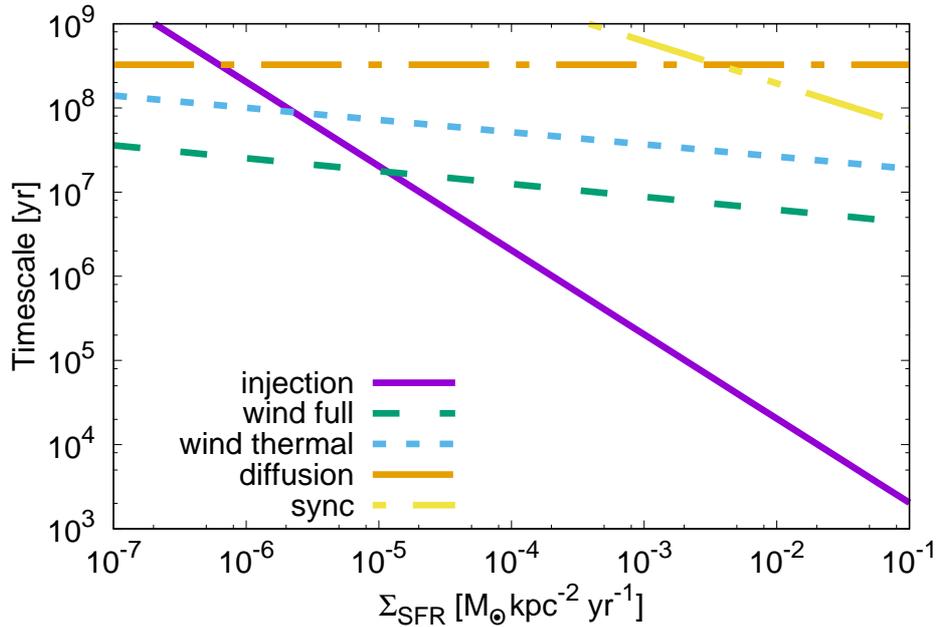}
\caption{Characteristic timescales for cosmic ray loss mechanisms for a reference model with $R_{\rm disk}=0.5$~kpc and a rotational velocity of $20$~km/s of the star-forming region. Shown are in particular the injection timescale of the cosmic rays, defined as the timescale for massive star formation, the adiabatic losses both for a full and a thermally-driven wind model, the timescale for cosmic ray diffusion and for synchrotron losses. The dominant loss mechanism in this regime is due to the full wind model, while cosmic ray diffusion would imply a transition for lower star formation rates. Injection timescales longer than the characteristic timescales for losses may induce significant fluctuations in the non-thermal radio emission. We note that the cosmic ray diffusion timescale implicitly assumes an observed frequency of $1$~GHz, while higher-frequency observations may probe more energetic cosmic rays with shorter diffusion times.}
\label{timescale20}
\end{center}
\end{figure*}

\begin{figure*}[htbp]
\begin{center}
\includegraphics[scale=1]{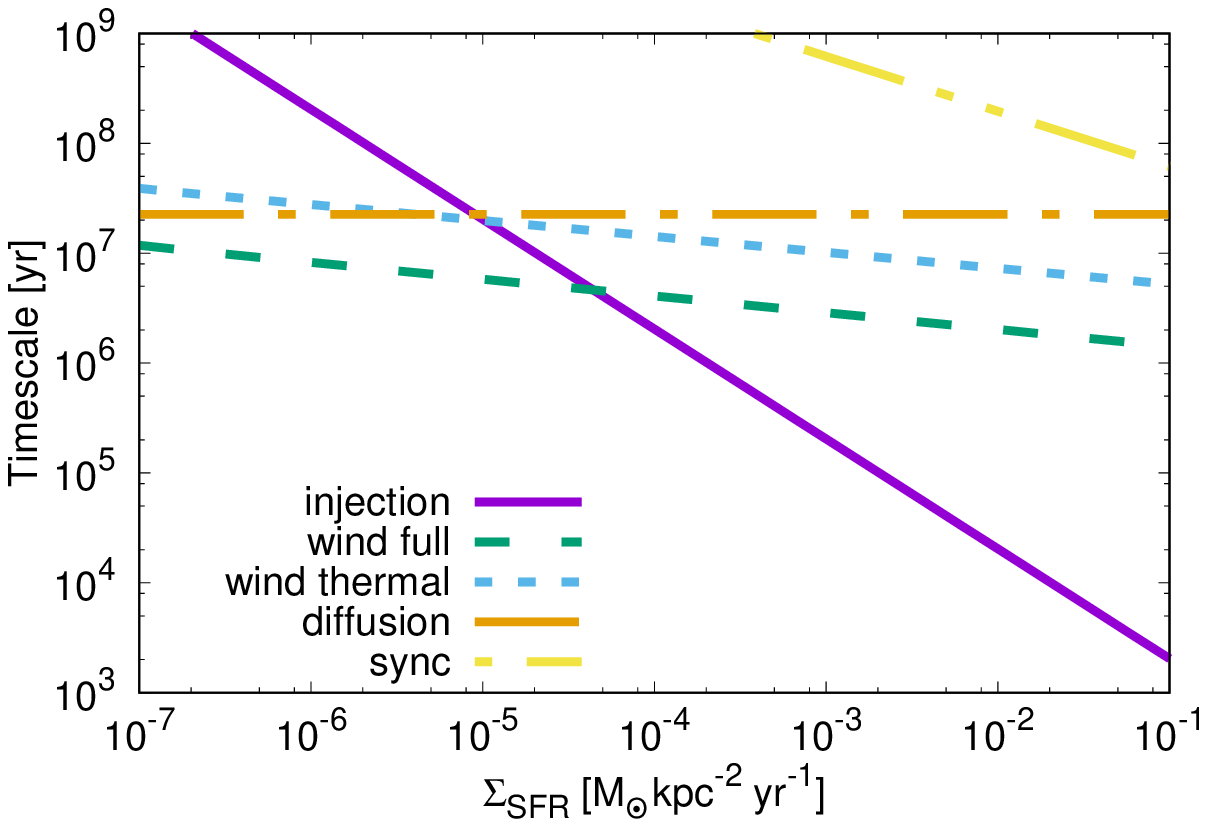}
\caption{{Same as Fig.~\ref{timescale20}, but with a rotational velocity of $40$~km/s.}}
\label{timescale40}
\end{center}
\end{figure*}

\begin{figure*}[htbp]
\begin{center}
\includegraphics[scale=1]{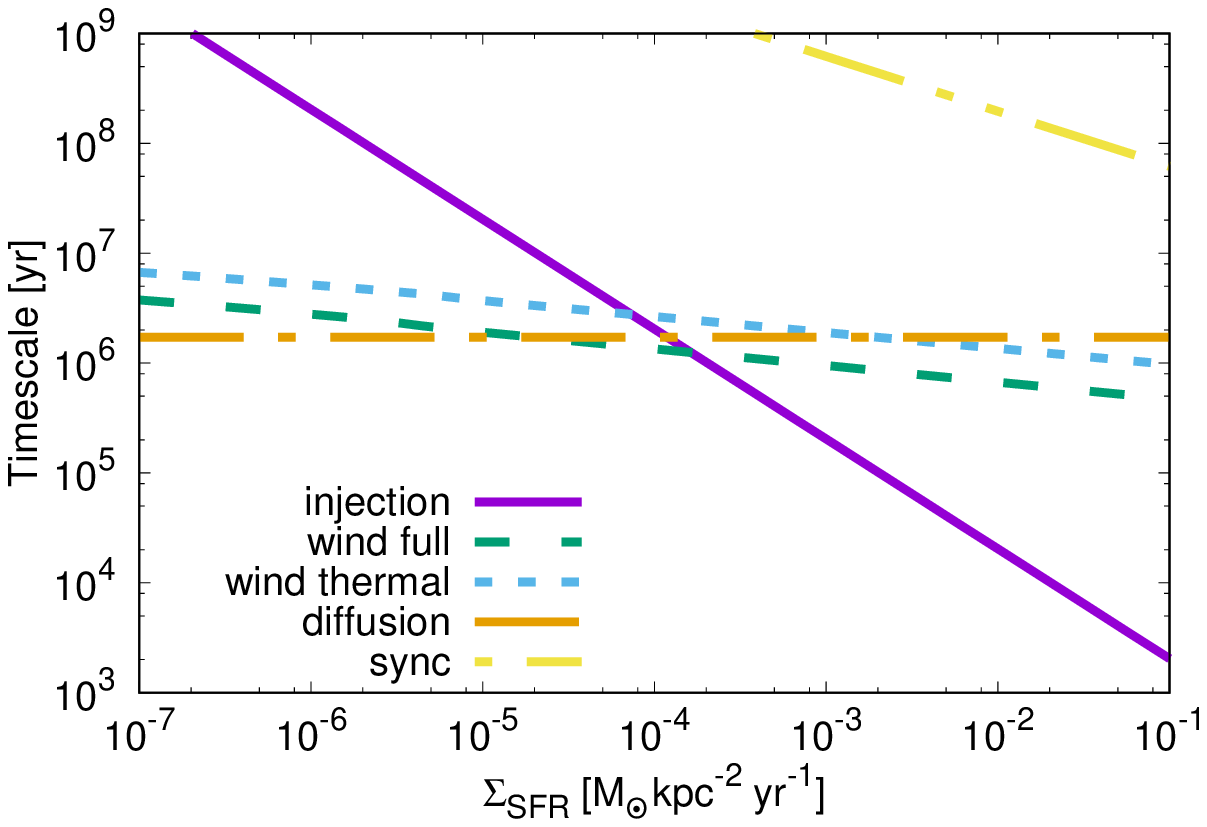}
\caption{{Same as Fig.~\ref{timescale20}, but with a rotational velocity of $80$~km/s.}}
\label{timescale80}
\end{center}
\end{figure*}

The injection of turbulent energy is closely linked to the formation of massive stars with at least $8$~M$_\odot$, which have a typical lifetime of  about $55$~million years. The formation rate of massive stars is given as\begin{equation} 
\dot{M}_{\rm sf,hm}=\epsilon_{hm}\dot{M}_{\rm SFR}=\epsilon_{hm}R_{\rm disk}^2\pi C\Sigma^N,
\end{equation}
with $\epsilon_{hm}$ the mass fraction of massive stars introduced above. {The characteristic timescale between the formation of two massive stars is then given as}\begin{equation}
\tau_{\rm hm}=\frac{8\ M_\odot}{\dot{M}_{\rm sf,hm}}=\frac{8\ M_\odot}{\epsilon_{hm}R_{\rm disk}^2\pi C\Sigma^N},\label{sfhm}
\end{equation}
corresponding to the characteristic timescale of turbulent energy injection. {The resulting scaling with $\Sigma_{\rm SFR}^{-1}\propto\Sigma^{-N}$ is directly visible in Fig.~\ref{sfr_diss}.} Now, a sufficient amount of turbulence in the galaxy can be maintained as long as $\tau_{\rm hm}<\tau_{\rm diss}$, implying that the turbulent energy is efficiently replenished by star formation. In the opposite case with $\tau_{\rm hm}>\tau_{\rm diss}$, the turbulence in the galaxy will decay before the next injection event. The latter implies the decay of turbulence and the decay of magnetic fields, implying that a relation such as $B\propto \Sigma_{\rm SFR}^{1/3}$ cannot be maintained. In particular, one may expect a significant increase of the scatter depending on the state at which the dwarf galaxy will be observed.

From this condition, one can derive a critical gas surface density above which the turbulence injection timescale remains sufficiently small:\begin{equation}
\Sigma_{\rm crit}=\left( \frac{8\ M_\odot\Omega_K}{\epsilon_{hm}R_{\rm disk}^2\pi C} \right)^{1/N}.
\end{equation}
Using the Kennicutt-Schmidt-relation (Eq.~\ref{Kennicutt}), the latter can be translated into a critical star formation surface density, yielding\begin{equation}
\Sigma_{\rm SFR, crit}=\frac{8\ M_\odot\Omega_K}{\epsilon_{hm}R_{\rm disk}^2\pi}.\label{SFRcrit}
\end{equation}

For a prelimary estimate, we adopt here a radius of $R_{\rm disk}=0.5$~kpc for the size of the star forming region, and we investigate the turbulence injection and dissipation timescales for dwarf galaxies with rotational velocities of $20$~km/s, $40$~km/s and $80$~km/s for the star-forming region. The critical star formation surface density is found to be $\sim10^{-5}$~M$_\odot$~kpc$^{-2}$~yr$^{-1}$, with only a weak dependence of the rotational velocity.  The more detailed results are given in Fig.~\ref{sfr_diss}. We further explore the impact of different power-law slopes $N$ in the Kennicutt-Schmidt relation, as shown in Fig.~\ref{sfr_var}, and find only a weak dependence on the form of the relation. The same is true for changes in the normalization of the relation by up to a factor of 10.

\subsubsection{Timescales for thermal emission}
Our considerations so far have predominantly concerned the dynamics in the galaxy and whether they can maintain a power-law relation between star formation and the magnetic field strength. However, to maintain the observed far-infrared - radio correlation, the star formation rate needs to be translated into thermal emission from dust grains, while the radio emission is due to cosmic ray synchrotron emission in the magnetic field. The infrared emission is due to thermal emission of dust grains, and the corresponding timescale is the cooling time of the dust. It can be shown that the latter is almost independent of the grain-size, and mostly depends on the dust temperature \citep{Krugel08}. Assuming a dust temperature of $10$~K, the cooling time corresponds to $\sim10^4$~yrs, and is even shorter for larger dust temperatures. The latter implies that the radiation of massive stars deposited onto dust grains is radiated away very quickly. However, the thermal energy of the grains is replenished during the lifetime of the massive stars, which is of the order $50\times10^6$~yrs for a star with $8$ solar masses.

Considering Figs.~\ref{sfr_diss} and \ref{sfr_var}, we find that the timescale for massive star formation becomes longer than the lifetime of a massive star for a star formation surface density of $\sim3\times10^{-5}$~M$_\odot$~kpc$^{-2}$~yr$^{-1}$, considering our reference model with $R_{\rm disk}=0.5$~kpc.  We can further generalize this requirement for arbitrary sizes of the star forming region, yielding a critical star formation surface density for the thermal emission as\begin{equation}
\left(\frac{\Sigma_{\rm SFR, therm}}{10^{-6}\,M_\odot\,kpc^{-2}\, yr^{-1}}\right)=\frac{8\ M_\odot}{5\times10^7\ \mathrm{yr} (\frac{\epsilon_{\rm hm}}{0.05})(\frac{R_{\rm disk}}{\rm kpc})^2 \pi }.\label{thermal}
\end{equation}
{For star formation surface densities lower than $10^{-6}$~M$_\odot$~kpc$^{}-2$~yr$^{-1}$, assuming a star-forming region of $1$~kpc and $\epsilon_{hm}\sim0.05$}, a constant thermal emission cannot be maintained, leading to strong fluctuations in the far-infrared - radio correlation. A steepening of the initial mass function (IMF) at very low star formation rates may somewhat change this transition, and in that case a breakdown of continuous thermal emission might occur even earlier. Understanding such an effect however requires a better understanding of star formation in this regime.

From a comparison with Eq.~\ref{SFRcrit}, we note that the condition becomes relevant for $\Omega_{K,therm}^{-1}<5\times10^7$~yr, as then the continuous thermal emission breaks down while the relation $B\propto\Sigma_{\rm SFR}^{1/3}$ is still maintained. In the opposite case, a breakdown of the relation between $B$ and $\Sigma_{\rm SFR}$ may occur while still having a steady thermal emission. The rotation rate of the galaxy is thus a key parameter in regulating this transition.

\subsubsection{Timescales for radio emission}\label{radioloss}
In the following, we will assess the characteristic timescales for radio emission and cosmic ray energy losses and their impact on the far-infrared - radio correlation. The timescale for the synchrotron emission can be expressed as \citep{Murphy09}\begin{equation}
\frac{\tau_{\rm sync}}{\rm yr}=1.4\times10^9\left( \frac{\nu_c}{\rm GHz} \right)^{-1/2}\left( \frac{B}{\mu G} \right)^{-3/2},
\end{equation}
where $\nu_c$ is the characteristic frequency of emission of cosmic rays with an energy $E$. These quantities are related via \citep{Murphy09}\begin{equation}
\frac{\nu_c}{\rm GHz}=1.3\times10^{-2}\left( \frac{B}{\mu G} \right)\left(\frac{E}{\rm GeV} \right)^2.\label{nuc}
\end{equation}
In the following, we will consider the synchrotron emission of cosmic rays with a characteristic frequency of $1$~GHz, and we evalute the magnetic field strength assuming the observed relation $B\propto\Sigma_{SFR}^{1/3}$ (Eq.~\ref{BSFR}). This timescale, along with other characteristic timescales introduced below, is shown in Figs.~\ref{timescale20}-\ref{timescale80} for dwarf galaxies with a star forming region of $R_{\rm disk}=0.5$~kpc and rotational velocities of $20$~km/s, $40$~km/s and $80$~km/s. In particular, the synchrotron timescale does not depend on the amount of rotation, it decreases with $\Sigma_{\rm SFR}$ and is considerably larger than the other timescales. We can therefore conclude that the cosmic ray abundance will not be significantly depleted via synchrotron emission.

An additional effect which can be relevant are the inverse Compton losses of the non-thermal electrons. Considering the inverse Compton scattering due to the cosmic microwave background (CMB), it can be shown that the latter becomes important at a critical field strength \citep{Murphy09, Schleicher13b} \begin{equation}
B_{IC}=3.25\,\mu\mathrm{G}\left(1+z\right)^2.\label{BIC}
\end{equation}
As the magnetic field strength $B$ scales approximately as $\Sigma_{\rm SFR}^{1/3}$ (cf. Eq.~\ref{BSFR}), we expect that inverse Compton losses become relevant in dwarf galaxies. Comparing the Eq.~(\ref{BSFR}) with Eq.~(\ref{BIC}), these losses are expected to become dominant for star formation surface densities below $0.005$~M$_\odot$~kpc$^{-2}$~yr$^{-1}$. In this regime, the expected synchrotron flux should thus be corrected by a factor $\tau_{IC}/\tau_{\rm sync}\propto B^2\propto \Sigma_{\rm SFR}^{2/3}$  \citep{Murphy09}, implying a steeper decrease of the non-thermal radio emission at very low star formation rates.

In the presence of further cosmic ray loss mechanisms with timescales shorter than the characteristic timescale for the injection, the cosmic rays will be depleted very efficiently, inducing an even steeper scaling relation, along with significant fluctuations in the non-thermal radio emission with peaks occuring at the events of cosmic ray injection. In particular, the cosmic rays will be depleted through diffusion processes in the interstellar medium.  Typical values of the diffusion coefficients range from $3\times10^{27}$~cm$^2$~s$^{-1}$ \citep{Mulcahy14} to $2\times10^{29}$~cm$^2$~s$^{-1}$ \citep{Heesen09}, and may depend both on the magnetic field strength and the alignment of the magnetic fields. We adopt here a typical diffusion coefficient of $D_E=2\times10^{28}$~cm$^2$~s$^{-1}$, which is characteristic for cosmic ray energies of $1$~GeV and which allows us to assess the potential relevance of the diffusion \citep{Murphy09}. The characteristic diffusion timescale of the cosmic rays is then given as\begin{equation}
\frac{\tau_{\rm esc}}{yr}=1.5\times10^7\left( \frac{H}{\rm kpc} \right)^2,
\end{equation}
where we assumed that the shortest pathway for the cosmic rays to diffuse out of the galaxy is along the disk height. Assuming observations at a fixed frequency of $1$~GHz, the energy of the cosmic rays providing the main contribution to the synchrotron emission is expected to weakly increase with decreasing magnetic field strength, potentially increasing the diffusion coefficient by a factor of 2-3. We however neglect that here in the light of the overall uncertainties. Another potential uncertainty in this timescale corresponds to the cosmic ray streaming instability \citep{McKenzie84}, as the typical Alfven time \begin{equation}
t_A\sim10^7\,\mathrm{yr}\left( \frac{H}{\mathrm{kpc}} \right)\left( \frac{v_A}{100\,\mathrm{km/s}} \right)^{-1}
\end{equation}
is comparable to the escape time by diffusion. The interaction of such processes thus should be treated in more detail via numerical simulations. 

With our simplified assumption, the diffusion timescale here depends on the rotational velocity, as the latter influences the scale height of the disk. For a rotational velocity of $20$~km/s, it is evident from Fig.~\ref{timescale20} that the diffusion timescale becomes shorter than the injection timescale at a star formation surface density of $\sim7\times10^{-7}$~M$_\odot$~kpc$^{-2}$~yr$^{-1}$. In case of a rotational velocity of $40$~km/s, this transition occurs at about $10^{-5}$~M$_\odot$~kpc$^{-2}$~yr$^{-1}$ (Fig.~\ref{timescale40}), and for a rotational velocity of $80$~km/s at about $10^{-4}$~M$_\odot$~kpc$^{-2}$~yr$^{-1}$ (Fig.~\ref{timescale80}). The rotational velocity therefore has a strong impact on the transition, predominantly as it implies a flatter structure potentially enhancing the escape. We note that the plots showing the diffusion timescale implicitly assume an observed frequency of $1$~GHz. Higher-frequency observations could probe more energetic cosmic rays, implying shorter diffusion times.

From a comparison with Eq.~\ref{sfhm}, we can derive a critical star formation surface density below which the cosmic ray diffusion becomes relevant. In this regime, the cosmic rays will be strongly depleted after the ejection event, implying that the radio emission cannot be maintained. The critical rate is given as\begin{equation}
\Sigma_{\rm SFR, diff}=\frac{8\ M_\odot}{1.5\times10^7\ \mathrm{yr}\times\epsilon_{hm}R_{\rm disk}^2\pi (H/\mathrm{kpc})^2},\label{SFRdiff}
\end{equation}
implying an increasing critical star formation rate of larger disk heights. Observationally the latter may be estimated from a measurement of the turbulent velocity $v_t$ and the rotation rate $\Omega_K$, thus yielding $H\sim v_t/\Omega_K$. We further note that this process will depend on the cosmic ray energy, as more energetic cosmic rays have larger diffusion coefficients, and longer diffusion times. We therefore expect that observations at higher frequencies will allow to correct for the effect of cosmic ray diffusion and probe whether a relation between the star formation rate and the magnetic field is still maintained. From a comparison of Eq.~\ref{SFRdiff} with Eq.~\ref{SFRcrit}, a critical point corresponds to the regime where $\Omega_K^{-1}=2.4\times10^6$~yr\ $(H/\mathrm{kpc})^2$. Using $H\sim v_t/\Omega_K$, the latter yields a critical transition at \begin{equation}
\Omega_{\rm K,diff}\sim1.7\times10^{-11} \left(\frac{v_t}{\mathrm{km/s}}\right)^2\ \mathrm{yr}^{-1},\label{omegadiff}
\end{equation}
or, focusing on the dependence of the scale height of the disk,\begin{equation}
\Omega_{\rm K,diff}=\frac{1}{1.5\times10^7\mathrm{\ yr}\times(H/\mathrm{kpc})^2}.\label{omegadiffdisk}
\end{equation}
For smaller angular velocities, the continuous radio emission may break down due to cosmic-ray diffusion even if the correlation between the magnetic field and the star formation rate is still maintained. This behavior can be inverted in the regime of fast rotation, where cosmic ray diffusion losses are expected to become relevant only after the breakdown of the correlation between the star formation rate and the magnetic field. We may further compare with the expression for the critical star formation rate for thermal emission (Eq.~\ref{thermal}), leading to the critical condition that\begin{equation}
\left( \frac{(v_t/\Omega_K)_{\rm crit}}{\rm kpc} \right)\sim1.8.
\end{equation}
In the regime of large disk heights, the thermal emission thus breaks down before the radio emission, while the inverse behavior is found for very thin disks.

Finally, we will consider adiabatic expansion losses due to galactic winds. The characteristic timescale can be estimated via \citep{Murphy09}\begin{equation}
\frac{\tau_{ad}}{yr}=10^9\left(\frac{H}{\rm kpc} \right)\left( \frac{v_{\rm wind}}{\rm km/s} \right)^{-1}.
\end{equation}
For the latter, we employ the detailed wind model outlined in subsection~\ref{windmodel}, and we will both consider the effects of thermally-driven winds (Eq.~\ref{windtherm}) as well as our full model including magnetic field and cosmic-ray pressure (Eq.~\ref{wind}). The effect of the wind model again has a strong dependence on the disk height of the galaxy, and thus its rotational velocity. In the reference case with a rotational velocity of $20$~km/s (Fig.~\ref{timescale20}), cosmic ray diffusion is less efficient than the effects of the thermal wind model, though comparable within an order of magnitude at the star formation surface density where it becomes comparable to the timescale for injection events. Including magnetic and cosmic ray pressure enhances the wind and implies a depletion of the cosmic rays already at a star formation rate of $\sim10^{-5}$~M$_\odot$~kpc$^{-2}$~yr$^{-1}$. For a rotational velocity of  $40$~km/s (Fig.~\ref{timescale40}), the effects of the thermal wind model and cosmic ray diffusion are essentially comparable, while magnetically driven winds would be even more efficient.  Finally at rotational velocities of  $80$~km/s (Fig.~\ref{timescale80}),cosmic ray diffusion is comparable to the effects of the full wind model.

Our results thus suggest that adiabatic losses by the wind can be particularly relevant in the regime of low rotation and large disk heights. From these results, it appears that the presence of a magnetically driven wind could change the critical star formation rate for cosmic ray depletion by at most an order of magnitude.

\subsubsection{Critical star formation surface densities}\label{criticalsfr}
\begin{figure*}[htbp]
\begin{center}
\includegraphics[scale=1]{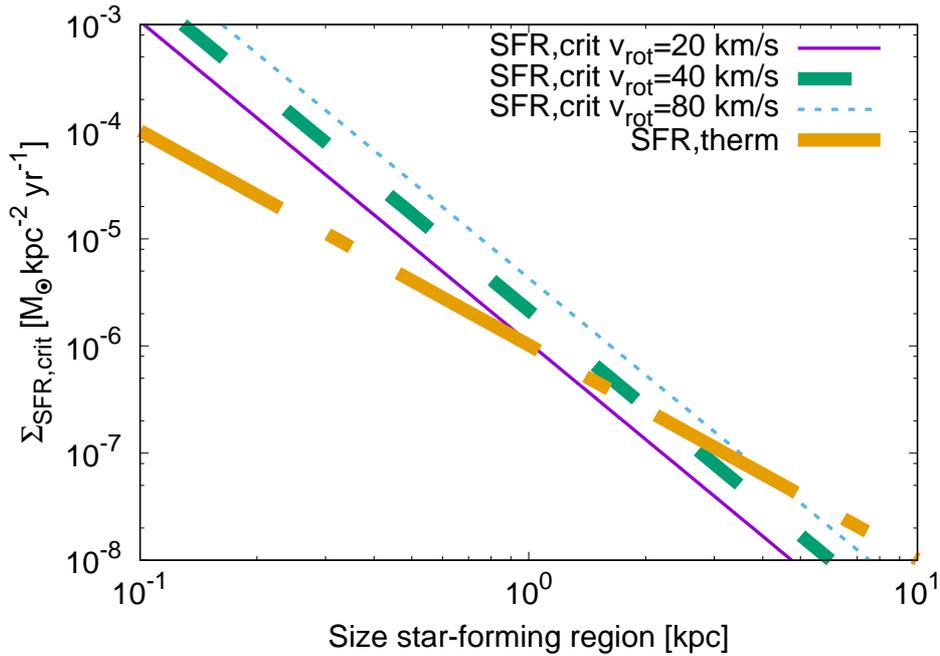}
\caption{A comparison of the critical star formation surface density to maintain the relation between star formation and the magnetic field, given via Eq.~\ref{SFRcrit}, with the critical star formation rate to maintain thermal emission, given via Eq.~\ref{thermal}. {Critical star formation surface densities are given as functions of the size of the star-forming region.} Explored are characteristic parameters for the rotational velocity of $20, 40, 80$~km/s. For large star-forming regions, the critical star formation rate to maintain thermal emission eventually exceeds the critical star formation rate to maintain the correlation between star formation and the magnetic field. A comparison with Fig.~\ref{sfrdiff} however shows that the radio emission will break down even earlier in this regime due to cosmic-ray diffusion losses.}
\label{sfrcrit}
\end{center}
\end{figure*}

\begin{figure*}[htbp]
\begin{center}
\includegraphics[scale=1]{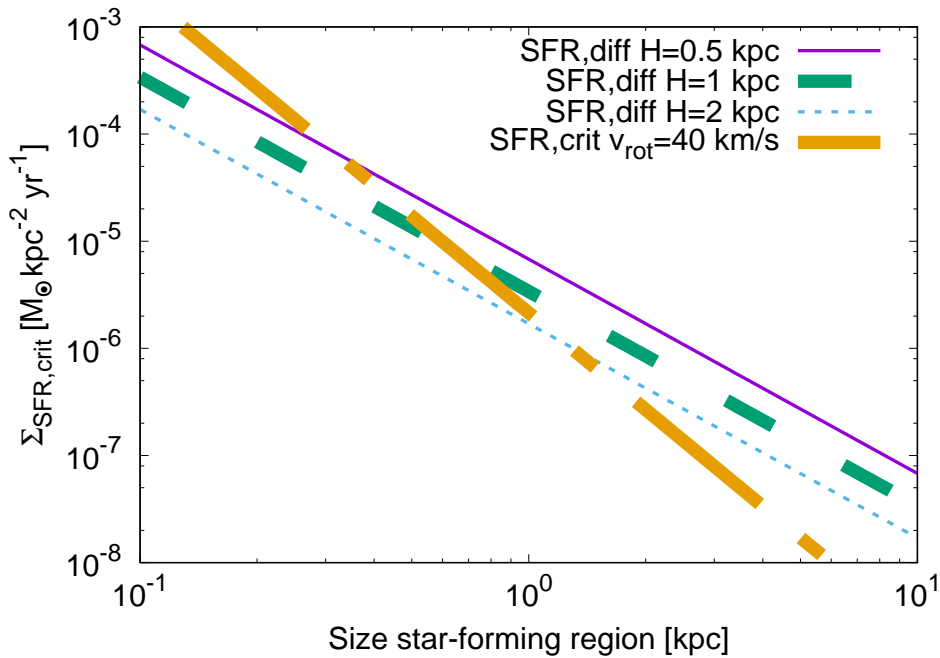}
\caption{A comparison of the critical star formation surface density for cosmic ray diffusion, given via Eq.~\ref{SFRdiff}, with the critical star formation rate to maintain the relation between star formation and magnetic field. {Critical star formation surface densities are given as functions of the size of the star-forming region.} We adopt here a characteristic rotational velocity of $20, 40, 80$~km/s, and explore values of the  disk scale height of $0.5, 1, 2$~kpc. The largest critical values can be obtained for more compact star-forming regions. The relative importance of both processes is found to sensitively depend on both quantities. We note that the critical star formation rate for cosmic ray diffusion implicitly assumes observations at $1$~GHz, while higher-frequency observations will probe more energetic cosmic rays, potentially implying a larger critical star formation rate.}
\label{sfrdiff}
\end{center}
\end{figure*}

From the considerations above, we have derived a set of critical star formation surface densities which describe the transition between different physical regimes in the dwarf galaxy and its ISM. Equation~(\ref{SFRcrit}) describes the {condition} under which the turbulence injection time remains larger than its decay time, implying that turbulence and turbulent magnetic field structures can be maintained and will be correlated with the star formation rate in the galaxy. This timescale depends in particular on the size of the star-forming region $R_{\rm disk}$ and the angular velocity $\Omega_K$, which can be expressed as $v_{\rm rot}/R_{\rm disk}$. 

Equation~(\ref{thermal}) denotes the critical star formation rate above which continuous thermal emission can be maintained in the galaxy, as the characteristic timescale for massive star formation is shorter than the typical lifetime of massive stars. This timescale depends essentially on the size of the star-forming region. Finally, equation~(\ref{SFRdiff}) denotes the critical star formation surface density for which the timescale for diffusion losses becomes comparable to the injection timescale of cosmic rays. This condition is therefore important to allow for continuous radio emission. The critical rate depends in particular on the size of the star-forming region and the disk height $H$.

\begin{figure*}[htbp]
\begin{center}
\includegraphics[scale=1]{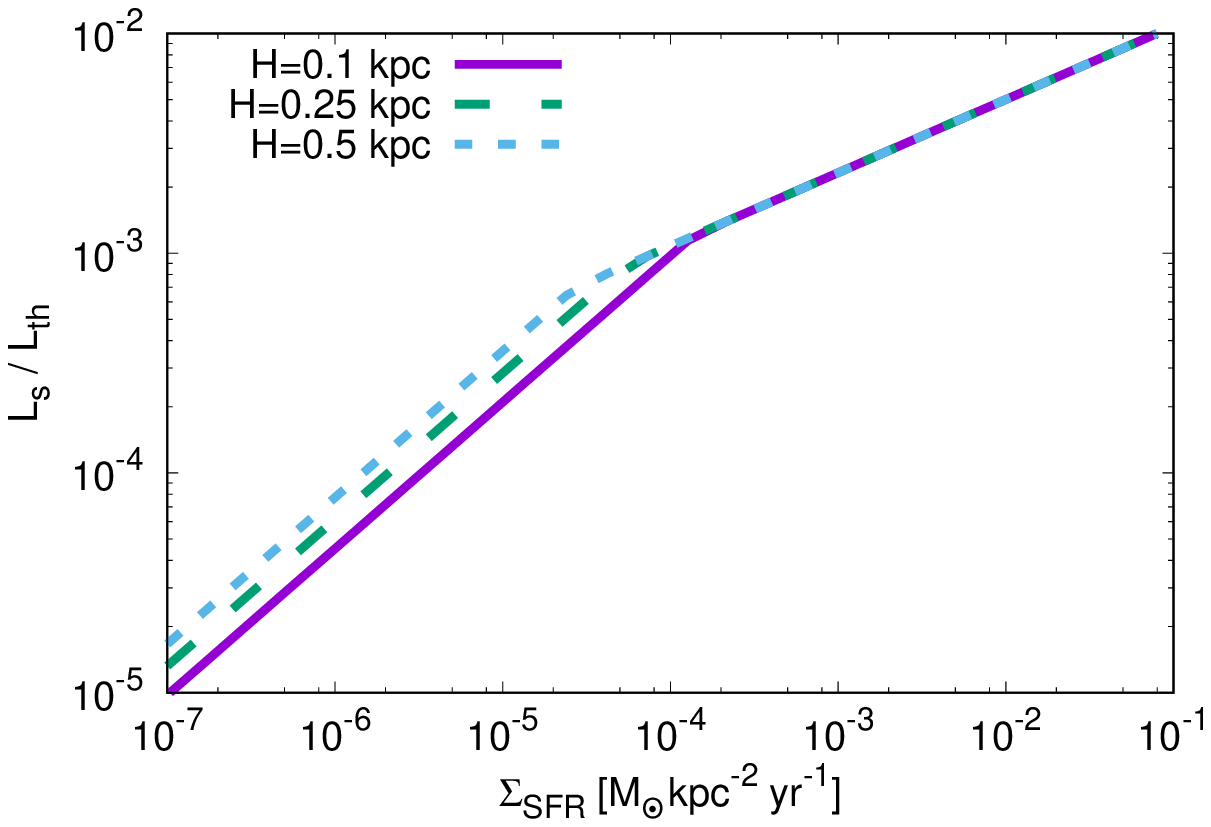}
\caption{Ratio of the non-thermal radio ($\sim2.64$~GHz) to thermal infrared ($\sim60$~$\mu$m) surface brightness, normalized for NGC~4449 adopting the values of \citet{Chyzy11}, as a function of the star formation surface density. We assume a size of the star-forming region of $0.5$~kpc, and explore values for the disk height of $0.1, 0.25, 0.5$~kpc, which influence the critical star formation surface density concerning the relevance of cosmic-ray diffusion losses (Eq.~\ref{SFRdiff}). The transition illustrated here is relevant in the regime of slow rotation defined via Eq.~(\ref{omegadiffdisk}), {where cosmic-ray diffusion effects become relevant before the relation between magnetic field strengths and star formation surface density breaks down}. }
\label{lum}
\end{center}
\end{figure*}

\begin{figure*}[htbp]
\begin{center}
\includegraphics[scale=1]{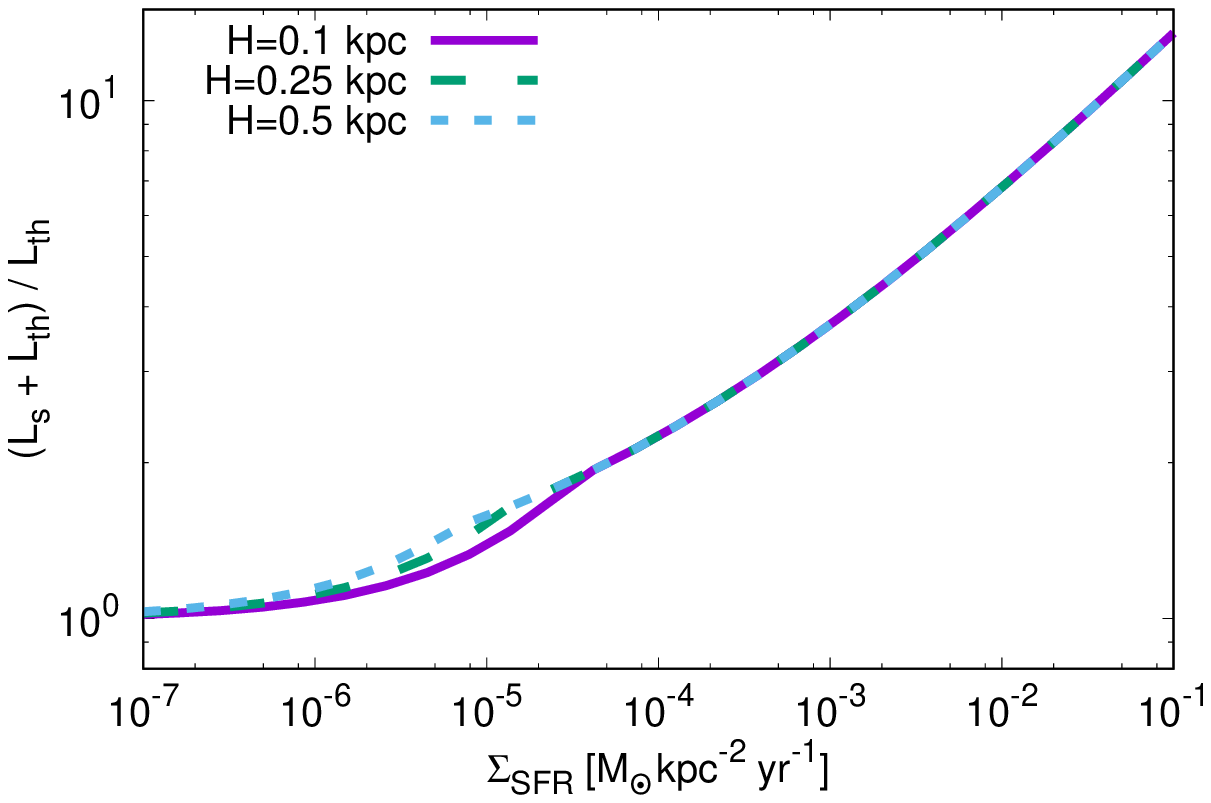}
\caption{The total to thermal radio flux density ratio as a function of the star formation rate. The evolution of the slope is eveluated from Eqs.~(\ref{spiral}, \ref{redCR}), assuming a transition at the critical star formation rate given via Eq.~\ref{SFRdiff}. We assume a size of the star-forming region of $0.5$~kpc, and explore values for the disk height of $0.1, 0.25, 0.5$~kpc. The normalization of the flux ratios assumes a thermal-to-non-thermal flux ratio of $8\%$ for a star formation rate of $0.1$~M$_\odot$~kpc$^{-2}$~yr$^{-1}$. The plot assumes a typical radio frequency of $\sim1$~GHz.}
\label{lum_therm}
\end{center}
\end{figure*}

To assess the relative importance of the different critical star formation surface densities, they are plotted as a function of $R_{\rm disk}$ in Fig.~\ref{sfrcrit}-\ref{sfrdiff}, considering typical rotational velocities $v_{\rm rot}=20, 40, 80$~ km/s and disk scale heights $H=0.5, 1, 2$~kpc. In particular for large star-forming regions, the dominant mechanism leading to a break-down of the far-infrared - radio correlation is cosmic ray diffusion, however with critical values of $\sim10^{-6}$~M$_\odot$~kpc$^{-2}$~yr$^{-1}$, which are difficult to observe. The critical values are significantly enhanced for more compact star-forming regions, scaling approximately as $R_{\rm disk}^{-2}$, with characteristic values of about  $\sim10^{-4}$~M$_\odot$~kpc$^{-2}$~yr$^{-1}$ for $R_{\rm disk}=0.5$~kpc. In this case, the critical values due to cosmic ray diffusion and due to the comparison of turbulence injection and dissipation time are about equally important. The critical star formation rate to maintain thermal emission, on the other hand, never appears to be responsible for a critical transition, as the lifetime of massive stars is still long enough for other processes to become relevant earlier. 

We note again that the critical star formation surface density for cosmic ray diffusion implicitly assumes observations at a frequency of $1$~GHz. Observations at higher frequencies will probe more energetic cosmic rays and imply shorter timescales for cosmic ray diffusion, potentially increasing the critical star formation rate to maintain the injection events at a high enough rate.

\section{The slope of the far-infrared - radio correlation}\label{slope}
The synchrotron emission from an ensemble of relativistic electrons of energy $E$ at a frequency $\nu$ is generally given as\begin{equation}
L_s(\nu)d\nu=\frac{dE}{dt}N(E)dE,
\end{equation}
where $dE/dt$ describes the synchrotron emission of the individual electrons and $N(E)$ denotes the number of electrons at energy $E$. The losses of the individual electrons are directly proportional to the magnetic energy density $U_B=B^2/8\pi$.

In spiral galaxies, the injection timescale of cosmic rays is considerably smaller than the timescale for synchrotron losses, cosmic ray diffusion, losses by winds or other effects like inverse Compton scattering \citep[see e.g.][]{Murphy09}. In this regime, the amount of cosmic rays that can be maintained in the galaxy corresponds to equipartition with the magnetic field, implying $U_B=U_{CR}$, as for larger amounts of cosmic rays, the cosmic ray pressure will exceed the pressure from the magnetic field, leading to expansion and an effective loss of the cosmic rays from the galaxy. In this regime, we thus have $N(E)\propto B^2/8\pi$, and thus $L(\nu)\propto B^4\propto \Sigma_{\rm SFR}^{4/3}$, where we assumed that $B\propto\Sigma_{\rm SFR}^{1/3}$. In the more general case with a synchrotron spectral index $\alpha$ different from $1$, one may expect a scaling as $B^{3+\alpha}$, while we note that the currently observed value $\alpha=0.7\pm0.2$ is consistent with the above relation in the $2\Sigma$ range. The ratio of non-thermal to thermal emission is therefore expected to scale as \begin{equation}
\left(\frac{L_s}{L_{th}}\right)_{\rm spiral}\propto \Sigma_{\rm SFR}^{1/3},\label{spiral}
\end{equation}
as the thermal emission is expected to be proportional to the star formation rate. In this derivation, we have so far assumed that synchrotron emission is the main loss mechanism for the cosmic rays, and we assumed that galaxies are optically thick to the UV emission of the stars, implying that the energy is trapped and converted to thermal emission of the dust grains. As we have shown in section~\ref{radioloss}, inverse Compton losses may start to become relevant at star formation surface densities of $\sim0.005$~M$_\odot$~kpc$^{-2}$~yr$^{-1}$, leading to an additional factor $\Sigma_{\rm SFR}^{2/3}$ in the scaling of the non-thermal radio emission at very low star formation rates. 

To take into account the potential effect of optically thin dust opacities for the UV, we adopt here a typical UV dust opacity of $\kappa_d\sim300$~cm$^2~g^{-1}$ \citep{Tielens05}. The effects of the optically thin regime thus start to become relevant for surface densities of $\sim1.5\times10^7$~M$_\odot$~kpc$^{-2}$. From the Kennicutt-Schmidt relation (Eq.~\ref{Kennicutt}) with $N\sim3/2$, the latter corresponds to a star formation surface density of $\sim0.002$~M$_\odot$~kpc$^{-2}$~yr$^{-1}$. Within the uncertainties present here, this correction thus becomes relevant in the same regime in which the inverse Compton losses also become relevant. To correct for this effect, the thermal emission should be multiplied with a factor proportional to the dust UV opacity $\tau_{d}\propto \kappa_d \Sigma\propto \kappa_d \Sigma_{\rm SFR}^{1/N}$. In the typical case of $N=3/2$, we thus have a correction factor of $\Sigma_{\rm SFR}^{2/3}$, and the corrections from both effects essentially cancel, so we expect that Eq~(\ref{spiral}) is still valid in this regime. Even for the extreme cases $N=1$ or $N=2$, we note that the corrections would be small, corresponding to a dependence of the form $\Sigma_{\rm SFR}^{\pm1/6}$, i.e. a very weak change in the dependence on $\Sigma_{\rm SFR}.$ We note that this canceling can also be understood in terms of a decreasing efficiency of star formation tracers, as previously described by \citet{Bell03}.

{The scaling relation given in Eq.~(\ref{spiral})} is both expected (and observed) for the ratio of non-thermal to thermal radio emission, as well as for the ratio of non-thermal radio emission to thermal emission in the infrared or far-infrared, as the thermal component is always proportional to the star formation rate. The latter implies a non-linear far-infrared - radio correlation with a logarithmic slope of $4/3$, i.e. $L_s\propto L_{th}^{4/3}$ \citep{Niklas97b}. From the shape of Eq.~(\ref{spiral}), it is further evident that the ratio of non-thermal to thermal emission decreases with decreasing star formation rates. This is consistent with recent observations by \citet{Roychowdhury12} of non-thermal radio fractions of up to $50\%$ in dwarf galaxies, in contrast to typical thermal fractions of $\sim8\%$ in spirals \citep{Niklas97, Murphy06}. Potential deviations from such a trend could potentially occur due to a steepening of the IMF at very low star formation rates, which would reduce the thermal emission, but also the injection of the cosmic rays. We however assume here that such effects do not yet occur in the regime where equipartition between cosmic rays and magnetic fields can be efficiently maintained.

\begin{figure*}[htbp]
\begin{center}
\includegraphics[scale=1]{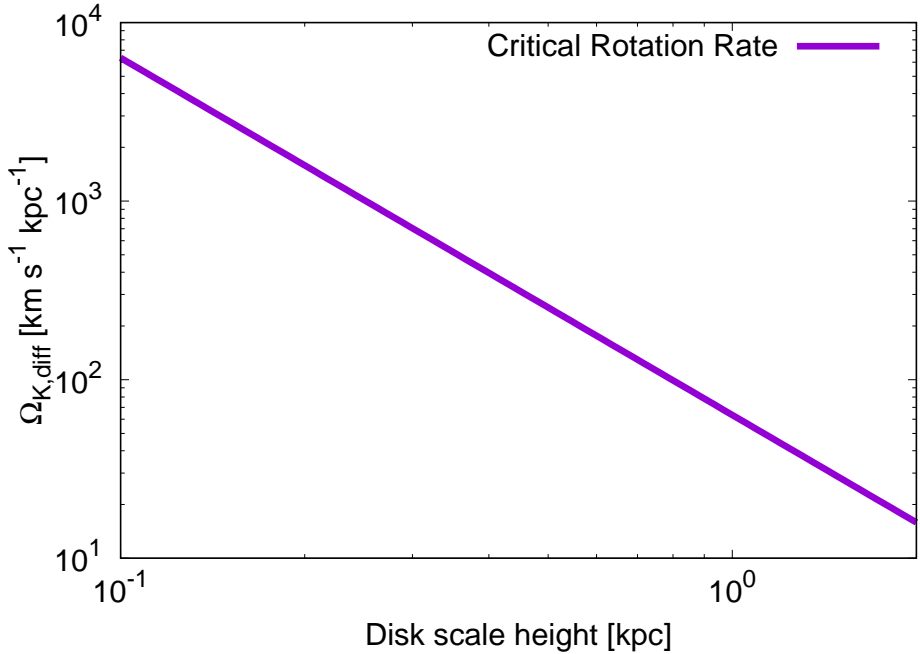}
\caption{The critical rotation rate  $\Omega_{\rm K, diff}$ of the dwarf galaxy as a function of the disk scale height $H$, as given in Eq.~(\ref{omegadiffdisk}), to distinguish between the regime of low star formation and high rotation and the regime of low star formation and low rotation. In the former regime, we still expect a valid but steeper far-infrared - radio correlation, while in the limit of low rotation, the correlation between star formation and the magnetic field will break down before cosmic ray diffusion becomes relevant. The critical rotation rate adopted here implicitly assumes observations at $1$~GHz due to the diffusion timescale of the cosmic rays. Observations at higher frequencies will probe more energetic electrons, implying shorter diffusion times, and a larger critical rotation rate.}
\label{omegacrit}
\end{center}
\end{figure*}

While a change of the thermal fraction may already be expected on the grounds given above, we will further illustrate below that also the slope of the far-infrared - radio correlation, or correspondingly, the slope of the thermal to non-thermal radio emission, will change with decreasing star formation rate. For this purpose, we need to consider the regime where cosmic ray injection becomes less efficient, so that a state of equipartition with the magnetic field is not reached. In particular, when the timescale for cosmic ray losses via diffusion or galactic winds becomes shorter than the injection timescale, i.e. the formation timescale of massive stars, the cosmic ray abundance in the galaxy will start decreasing. As we have shown in section~\ref{radioloss} that cosmic ray diffusion and adiabatic losses through winds are yielding roughly similar results, we will in the following focus on losses via cosmic ray diffusion, which are relevant below the critical star formation surface density given in Eq.~(\ref{SFRdiff}), with a strong dependence both on the size of the star-forming region $R_{\rm disk}$ and the disk scale height $H$. 

In this regime, the amount of cosmic rays will be dictated by the injection, implying that {their number $N(E)$ scales as} $N(E)\propto \Sigma_{\rm SFR}$. In this case, the ratio of non-thermal to thermal emission scales as
\begin{equation}
\left(\frac{L_s}{L_{th}}\right)_{\rm CR\ diff}\propto \frac{U_B \Sigma_{\rm SFR}}{\Sigma_{\rm SFR}}\propto B^2.\label{redCR}
\end{equation}
Under these conditions, the magnetic field strength $B$ depends on the rotation rate of the galaxy. As shown in section~\ref{radioloss}, there is a critical rotation rate $\Omega_{\rm K, diff}$ (Eq.~\ref{omegadiffdisk}) above which the relation between magnetic field and star formation rate is still maintained, while this relation is no longer valid for lower values of the rotation rate. In terms of the critical star formation rates derived above, one may thus distinguish between the following regimes:\begin{itemize}
\item high star formation rates with $\Sigma_{\rm SFR}> \Sigma_{\rm SFR,crit}$ and  $\Sigma_{\rm SFR}> \Sigma_{\rm SFR,diff}$,
\item low star formation rates and low rotation, $\Sigma_{\rm SFR}> \Sigma_{\rm SFR,crit}$ but  $\Sigma_{\rm SFR}< \Sigma_{\rm SFR,diff}$,
\item low star formation rates and strong rotation, $\Sigma_{\rm SFR}< \Sigma_{\rm SFR,crit}$ and  $\Sigma_{\rm SFR}> \Sigma_{\rm SFR,diff}$,
\item very low star formation rates with $\Sigma_{\rm SFR}< \Sigma_{\rm SFR,crit}$ and  $\Sigma_{\rm SFR}< \Sigma_{\rm SFR,diff}$,
\item very low star formation rates with $\Sigma_{\rm SFR}<\Sigma_{\rm SFR,therm}$.
\end{itemize}
The first case is the regime of spiral galaxies, which we described in Eq.~(\ref{spiral}). In the regime of low star formation and low rotation, the ratio of non-thermal to thermal emission can be described via Eq.~(\ref{redCR}), and a correlation $B\propto\Sigma_{\rm SFR}^{1/3}$ can still be expected. In this case, we have\begin{equation}
\left(\frac{L_s}{L_{th}}\right)_{\rm low\ SF,\ high\ rot}\propto B^2\propto \Sigma_{\rm SFR}^{2/3}, \label{lowSFhrot}
\end{equation}
implying a scaling of $L_s\propto L_{th}^{5/3}$ and a change in the slope of the far-infrared - radio correlation by $1/3$ on a logarithmic scale. In this regime, the ratio thus decreases more steeply with $\Sigma_{\rm SFR}$, and the thermal fraction will be further enhanced towards low star formation rates. The latter is consistent with the current stacking results by \citet{Roychowdhury12}.

 The latter is given in Fig.~\ref{lum} for a characteristic case with $R_{\rm disk}=0.5$~kpc and disk scale heights of $0.1, 0.25, 0.5$~kpc, with a normalization of the relation obtained via NGC~4449, {as the latter lies well onto the $L_{2.64\ GHz}-L_{60\ \mu m}$ relation given by \citet{Chyzy11}. While our results related to cosmic ray diffusion have been arrived assuming a frequency of $1$~GHz, we note that the characteristic electron energy scales as $E\propto\nu_c^{1/2}$, and the cosmic ray diffusion coefficient has a weak scaling $\propto E^{1/2}$. Overall, the difference in frequency thus introduces only a factor of $1.3$ difference with respect to cosmic ray diffusion. We also note that our results have a predominantly illustrative character, and the typical scatter in this relations needs to be taken into account, implying a variation by at least a factor of $3$.} 
 
{With the assumptions adopted here,}  the steepening occurs at a star formation surface density of $\sim3\times10^{-5}$~M$_\odot$~kpc$^{-2}$~yr$^{-1}$, with some spread depending on the scale height of the disk. We also show the ratio of the total to thermal radio emission in Fig.~\ref{lum_therm}. For normalization purposes, we have adopted a characteristic value of $8\%$ for the ratio of thermal to non-thermal radio emission at a star formation surface density of $0.1$~M$_\odot$~kpc$^{-2}$~yr$^{-1}$ \citep[see][for the thermal fraction in spiral galaxies]{Murphy06}. The figure shows a substantial decrease of the total to thermal emission as a function of the star formation rate, with a flattening towards a value of $1$ at very low star formation rates. The slope of this ratio changes at the critical star formation rate given in Eq.~(\ref{omegadiffdisk}), and the details of the transition depend on the properties of the dwarf galaxy.  As the distinction between this and the subsequent regime of low rotation depends on the critical rotation rate, the latter quantity is given in Fig.~\ref{omegacrit}.

 We note that the star formation rates for which we expect changes in the slope may also depend on the observed frequency, as cosmic ray diffusion becomes more efficient at larger observed frequencies, implying shorter diffusion times. High-frequency observations may thus be able to probe this transition already at higher star formation rates. At the same time, we note that the normalization of the thermal-to-non-thermal flux density ratio may be frequency dependent, and may require further exploration.

In the regime of low star formation rates and high rotation, there is no longer a correlation between star formation rate and magnetic field strength. The ratio of non-thermal to thermal emission can still be described by Eq.~(\ref{redCR}), but the magnetic field strength may be significantly decreased as a result of turbulent decay. In general, the latter is expected to significantly decrease the radio emission, while in a few cases with very recent injection, the non-thermal emission can be comparable to the case of low star formation and high rotational velocities. Quite generally, we expect here a significant amount of scatter, including many non-detections, and potentially a few number of detections of non-thermal emission.

In the case with very low star formation rates below $\Sigma_{\rm SFR,crit}$ and $\Sigma_{\rm SFR}< \Sigma_{\rm SFR,diff}$, the correlation between star formation and magnetic field is broken, and cosmic ray diffusion is very efficient, rapidly removing any cosmic rays that might be ejected. In this regime, we do not expect a relevant or detectable non-thermal component, and no signs of a far-infrared - radio correlation should be detectable. We expect that the corresponding dwarfs can be detected only via thermal emission. Continuous thermal emission is expected to eventually break down below $\Sigma_{\rm SFR}< \Sigma_{\rm SFR,therm}$.

\section{Discussion and conclusions}\label{discussions}
In this paper, we have discussed the far-infrared - radio correlation in dwarf galaxies and its potential evolution in the regime of low star formation rates. As a starting point, we have assessed under which conditions a correlation between the magnetic field strength and the star formation rate can be maintained through the continuous injection of turbulence, leading to the definition of a critical star formation rate required for the injection of a sufficient amount of turbulent energy (Eq.~\ref{SFRcrit}).  The latter will ensure magnetic field amplification via the small-scale dynamo \citep[e.g.][]{Kazantsev68, Scheko02, Schober12b}, which happens on short timescales and effectively ensures that the magnetic field strength is coupled to the star formation rate. {Considering a typical size of the star-forming region of $1$~kpc, this relation will break down at critical star formation surface densities of $10^{-5}-10^{-6}$~M$_\odot$~kpc$^{-2}$~yr$^{-1}$ depending on the amount of rotation.} We also derived a critical star formation rate to ensure continuous thermal emission in the galaxy, which results from the requirement that the timescale of massive star formation should be smaller than the typical lifetime of massive stars. This criterion has been developed in Eq.~\ref{thermal}, even though our comparison has shown that it is likely not relevant in practice. {Assuming a star-forming region of $1$~kpc and a $5\%$ fraction of massive stars, the resulting star formation surface density corresponds to $\sim10^{-6}$~M$_\odot$~kpc$^{-2}$~yr$^{-1}$. }

We further explored the dominant loss mechanisms for cosmic rays in the galaxy, which have a major impact on the non-thermal radio emission and its dependence on the star formation rate. In particular, we derived a critical star formation surface density for the importance of cosmic ray diffusion losses given by Eq.~(\ref{SFRdiff}), below which the injection time of cosmic rays becomes larger than the timescale for diffusion losses, thus strongly {reducing} the amount of cosmic rays in the galaxy. {Assuming a star-forming region of $1$~kpc, these effects become relevant for star formation surface densities of $10^{-4}-10^{-6}$~M$_\odot$~kpc$^{-2}$~yr$^{-1}$, depending on the scale height of the disk. Overall, the critical star formation surface density for this transition thus appears higher than the corresponding transition for the thermal emission. We have compared the cosmic-ray diffusion losses} with the losses due to cosmic winds, adopting both the thermally-driven wind model by \citet{Shu05} as well as a generalized model accounting for the effect of cosmic rays and magnetic fields as potential driving agents. Depending on the amount of rotation in the galaxy, we have found that the losses from winds can become relevant at somewhat higher or lower star formation rates compared to the cosmic ray diffusion processes, but do not change dramatically the details of the transition.

From a comparison of these results, we have shown that the critical star formation surface densities for the maintainance of turbulence (Eq.~\ref{SFRcrit}) and for the relevance of cosmic-ray diffusion (Eq.~\ref{SFRdiff}) are most relevant for introducing a potential breakdown in the far-infrared - radio correlation. Which of the transitions occurs first depends on the rotation rate of the dwarf and the scale height of the disk, and can be expressed in terms of a critical rotation rate given in Eq.~(\ref{omegadiffdisk}), {corresponding to rotation period of $1.5\times10^7$~yrs for a galaxy with scale height of $1$~kpc. We note in particular that cosmic-ray diffusion losses become relevant before the breakdown of the magnetic field - star formation surface density relation in the case of low rotation. Whether the rotation period is above or below the critical value thus regulates the further evolution of the far-infrared - radio correlation in the regime of low star formation rates.}

We have determined four different regimes employing the conditions derived above. In the limit of high star formation rates above the critical star formation surface densities {mentioned above}, we have shown that the ratio of non-thermal radio to far-infrared / infrared luminosity should scale as $\Sigma_{\rm SFR}^{1/3}$, as we expect that the cosmic ray abundance is limited by the strength of the magnetic field, effectively enforcing equipartition, implying a scaling of the cosmic-ray abundance with $B^2$ and of the overall non-thermal emission as $B^4$ or $\Sigma_{\rm SFR}^{4/3}$, {typically above a star formation surface density of $\sim10^{-6}$~M$_\odot$~kpc$^{-2}$~yr$^{-1}$ for a star-forming region of $1$~kpc.}

For lower star formation rates in the limit of low rotation (below the critical value in Eq.~\ref{omegadiffdisk}), the correlation between magnetic field strength and star formation will still be in place, but cosmic ray diffusion losses will start becoming efficient. As a result, the cosmic ray abundance in galaxies will no longer reach equipartition with the magnetic field strength, but is instead expected to be proportional to the injection rate, i.e. the rate of star formation. In this regime, it can then be shown that the ratio of non-thermal radio to far-infrared / infrared luminosity will scale as $\Sigma_{\rm SFR}^{2/3}$, implying a steeper decrease with decreasing star formation rate. Such steepening may be expected below typical star formation surface densities of $3\times10^{-5}$~M$_\odot$~kpc$^{-2}$~yr$^{-1}$, with some dependence both on the size of the star-forming region and the scale height of the disk. 

On the other hand, in the regime of low star formation rates and {rotation rates above the before-mentioned critical value}, the cosmic ray diffusion losses are not yet significant, but the correlation between the magnetic field strength and the star formation rate is expected to break down, as a steady injection of turbulence cannot be maintained, leading to the dissipation both of turbulence and the magnetic fields. As a result, one may expect significant non-thermal emission only in those sources with recent injection events, and otherwise no significant emission in the vast majority of sources. Finally, for star formation rates both below the critical rates for turbulence driving and cosmic-ray diffusion, it is clear that the correlation is broken down due to both mechanism.

Overall, we therefore expect modifications of the far-infrared - radio correlation in the regime of low star formation rates, which will be particularly relevant for future observations with the Square Kilometre Array (SKA)\footnote{Website SKA: https://www.skatelescope.org/} and its Key Science Project on ''The Origin and Evolution of Cosmic Magnetism''. Its precursors and pathfinders such as LOFAR\footnote{Website LOFAR: http://www.lofar.org/}, MeerKAT\footnote{Website MeerKAT: http://www.ska.ac.za/meerkat/} and ASKAP\footnote{Website ASKAP: http://www.atnf.csiro.au/projects/askap/} are in fact already in the process of further probing magnetic field structures in local dwarf galaxies, as in the LOFAR Key Science Project on ''Cosmic Magnetism of the Nearby Universe'' \citep{Beck13LOFAR}. Such dedicated investigations will be particularly valuable to determine the frontiers of cosmic magnetism and the origin of magnetic fields in galaxies.

\begin{acknowledgements}
We thank Chris Chyzy for valuable discussions on the rotation rates in dwarf galaxies, and Aritra Basu for valuable comments that helped to improve our manuscript. RB acknowledges support by the DFG Research Unit FOR1254. DRGS thanks for funding through Fondecyt regular (project code 1161247) and through the ''Concurso Proyectos Internacionales de Investigaci\'on, Convocatoria 2015'' (project code PII20150171).
\end{acknowledgements}


\begin{thebibliography}{88}
\expandafter\ifx\csname natexlab\endcsname\relax\def\natexlab#1{#1}\fi

\bibitem[{{Arshakian} {et~al.}(2009){Arshakian}, {Beck}, {Krause}, \&
  {Sokoloff}}]{Arshakian09}
{Arshakian}, T.~G., {Beck}, R., {Krause}, M., \& {Sokoloff}, D. 2009, A\&A,
  494, 21

\bibitem[{{Ashley} {et~al.}(2014){Ashley}, {Elmegreen}, {Johnson}, {Nidever},
  {Simpson}, \& {Pokhrel}}]{Ashley14}
{Ashley}, T., {Elmegreen}, B.~G., {Johnson}, M., {et~al.} 2014, \aj, 148, 130

\bibitem[{{Beck}(2007)}]{Beck07}
{Beck}, R. 2007, \aap, 470, 539

\bibitem[{{Beck}(2015)}]{Beck15}
{Beck}, R. 2015, \aap, 578, A93

\bibitem[{{Beck} {et~al.}(2013){Beck}, {Anderson}, {Heald}, {Horneffer},
  {Iacobelli}, {K{\"o}hler}, {Mulcahy}, {Pizzo}, {Scaife}, {Wucknitz}, \&
  {LOFAR Magnetism Key Science Project Team}}]{Beck13LOFAR}
{Beck}, R., {Anderson}, J., {Heald}, G., {et~al.} 2013, Astronomische
  Nachrichten, 334, 548

\bibitem[{{Begum} {et~al.}(2008){Begum}, {Chengalur}, {Karachentsev},
  {Sharina}, \& {Kaisin}}]{Begum08}
{Begum}, A., {Chengalur}, J.~N., {Karachentsev}, I.~D., {Sharina}, M.~E., \&
  {Kaisin}, S.~S. 2008, \mnras, 386, 1667

\bibitem[{{Bell}(2003)}]{Bell03}
{Bell}, E.~F. 2003, \apj, 586, 794

\bibitem[{{Bigiel} {et~al.}(2011){Bigiel}, {Leroy}, {Walter}, {Brinks}, {de
  Blok}, {Kramer}, {Rix}, {Schruba}, {Schuster}, {Usero}, \&
  {Wiesemeyer}}]{Bigiel11}
{Bigiel}, F., {Leroy}, A.~K., {Walter}, F., {et~al.} 2011, \apjl, 730, L13

\bibitem[{{Casey} {et~al.}(2012){Casey}, {Berta}, {B{\'e}thermin}, {Bock},
  {Bridge}, {Budynkiewicz}, {Burgarella}, {Chapin}, {Chapman}, {Clements},
  {Conley}, {Conselice}, {Cooray}, {Farrah}, {Hatziminaoglou}, {Ivison}, {le
  Floc'h}, {Lutz}, {Magdis}, {Magnelli}, {Oliver}, {Page}, {Pozzi},
  {Rigopoulou}, {Riguccini}, {Roseboom}, {Sanders}, {Scott}, {Seymour},
  {Valtchanov}, {Vieira}, {Viero}, \& {Wardlow}}]{Casey12}
{Casey}, C.~M., {Berta}, S., {B{\'e}thermin}, M., {et~al.} 2012, \apj, 761, 140

\bibitem[{{Chabrier}(2003)}]{Chabrier03}
{Chabrier}, G. 2003, \pasp, 115, 763

\bibitem[{{Chy{\.z}y} \& {Beck}(2004)}]{Chyzy04}
{Chy{\.z}y}, K.~T. \& {Beck}, R. 2004, \aap, 417, 541

\bibitem[{{Chy{\.z}y} {et~al.}(2000){Chy{\.z}y}, {Beck}, {Kohle}, {Klein}, \&
  {Urbanik}}]{Chyzy00}
{Chy{\.z}y}, K.~T., {Beck}, R., {Kohle}, S., {Klein}, U., \& {Urbanik}, M.
  2000, \aap, 355, 128

\bibitem[{{Chy{\.z}y} {et~al.}(2003){Chy{\.z}y}, {Knapik}, {Bomans}, {Klein},
  {Beck}, {Soida}, \& {Urbanik}}]{Chyzy03}
{Chy{\.z}y}, K.~T., {Knapik}, J., {Bomans}, D.~J., {et~al.} 2003, \aap, 405,
  513

\bibitem[{{Chy{\.z}y} {et~al.}(2011){Chy{\.z}y}, {We{\.z}gowiec}, {Beck}, \&
  {Bomans}}]{Chyzy11}
{Chy{\.z}y}, K.~T., {We{\.z}gowiec}, M., {Beck}, R., \& {Bomans}, D.~J. 2011,
  A\&A, 529, A94

\bibitem[{{Clarke} \& {Oey}(2002)}]{Clarke02}
{Clarke}, C. \& {Oey}, M.~S. 2002, \mnras, 337, 1299

\bibitem[{{Condon}(1992)}]{Condon92}
{Condon}, J.~J. 1992, \araa, 30, 575

\bibitem[{{Condon} {et~al.}(2002){Condon}, {Helou}, \& {Jarrett}}]{Condon02}
{Condon}, J.~J., {Helou}, G., \& {Jarrett}, T.~H. 2002, \aj, 123, 1881

\bibitem[{{de Jong} {et~al.}(1985){de Jong}, {Klein}, {Wielebinski}, \&
  {Wunderlich}}]{deJong85}
{de Jong}, T., {Klein}, U., {Wielebinski}, R., \& {Wunderlich}, E. 1985, \aap,
  147, L6

\bibitem[{{de Souza} \& {Opher}(2010)}]{Souza10}
{de Souza}, R.~S. \& {Opher}, R. 2010, \prd, 81, 067301

\bibitem[{{Donevski} \& {Prodanovi{\'c}}(2015)}]{Donevski15}
{Donevski}, D. \& {Prodanovi{\'c}}, T. 2015, \mnras, 453, 638

\bibitem[{{Drzazga} {et~al.}(2016){Drzazga}, {Chy{\.z}y}, {Heald}, {Elstner},
  \& {Gallagher}}]{Drzazga16}
{Drzazga}, R.~T., {Chy{\.z}y}, K.~T., {Heald}, G.~H., {Elstner}, D., \&
  {Gallagher}, J.~S. 2016, \aap, 589, A12

\bibitem[{{Drzazga} {et~al.}(2011){Drzazga}, {Chy{\.z}y}, {Jurusik}, \&
  {Wi{\'o}rkiewicz}}]{Drzazga11}
{Drzazga}, R.~T., {Chy{\.z}y}, K.~T., {Jurusik}, W., \& {Wi{\'o}rkiewicz}, K.
  2011, \aap, 533, A22

\bibitem[{{Dumas} {et~al.}(2011){Dumas}, {Schinnerer}, {Tabatabaei}, {Beck},
  {Velusamy}, \& {Murphy}}]{Dumas11}
{Dumas}, G., {Schinnerer}, E., {Tabatabaei}, F.~S., {et~al.} 2011, AJ, 141, 41

\bibitem[{{Efstathiou}(2000)}]{Efstathiou00}
{Efstathiou}, G. 2000, \mnras, 317, 697

\bibitem[{{Federrath} {et~al.}(2011){Federrath}, {Chabrier}, {Schober},
  {Banerjee}, {Klessen}, \& {Schleicher}}]{FederrathPRL}
{Federrath}, C., {Chabrier}, G., {Schober}, J., {et~al.} 2011, Physical Review
  Letters, 107, 114504

\bibitem[{{Gaensler} {et~al.}(2005){Gaensler}, {Haverkorn}, {Staveley-Smith},
  {Dickey}, {McClure-Griffiths}, {Dickel}, \& {Wolleben}}]{Gaensler05}
{Gaensler}, B.~M., {Haverkorn}, M., {Staveley-Smith}, L., {et~al.} 2005,
  Science, 307, 1610

\bibitem[{{Geha} {et~al.}(2013){Geha}, {Brown}, {Tumlinson}, {Kalirai},
  {Simon}, {Kirby}, {VandenBerg}, {Mu{\~n}oz}, {Avila}, {Guhathakurta}, \&
  {Ferguson}}]{Geha13}
{Geha}, M., {Brown}, T.~M., {Tumlinson}, J., {et~al.} 2013, \apj, 771, 29

\bibitem[{{Grete} {et~al.}(2015){Grete}, {Vlaykov}, {Schmidt}, {Schleicher}, \&
  {Federrath}}]{Grete15}
{Grete}, P., {Vlaykov}, D.~G., {Schmidt}, W., {Schleicher}, D.~R.~G., \&
  {Federrath}, C. 2015, New Journal of Physics, 17, 023070

\bibitem[{{Groves} {et~al.}(2003){Groves}, {Cho}, {Dopita}, \&
  {Lazarian}}]{Groves03}
{Groves}, B.~A., {Cho}, J., {Dopita}, M., \& {Lazarian}, A. 2003, \pasa, 20,
  252

\bibitem[{{Guo} {et~al.}(2014{\natexlab{a}}){Guo}, {Sironi}, \&
  {Narayan}}]{Guo14a}
{Guo}, X., {Sironi}, L., \& {Narayan}, R. 2014{\natexlab{a}}, \apj, 794, 153

\bibitem[{{Guo} {et~al.}(2014{\natexlab{b}}){Guo}, {Sironi}, \&
  {Narayan}}]{Guo14b}
{Guo}, X., {Sironi}, L., \& {Narayan}, R. 2014{\natexlab{b}}, \apj, 797, 47

\bibitem[{{Heesen} {et~al.}(2009){Heesen}, {Beck}, {Krause}, \&
  {Dettmar}}]{Heesen09}
{Heesen}, V., {Beck}, R., {Krause}, M., \& {Dettmar}, R.-J. 2009, Astronomische
  Nachrichten, 330, 1028

\bibitem[{{Heesen} {et~al.}(2014){Heesen}, {Brinks}, {Leroy}, {Heald}, {Braun},
  {Bigiel}, \& {Beck}}]{Heesen14}
{Heesen}, V., {Brinks}, E., {Leroy}, A.~K., {et~al.} 2014, \aj, 147, 103

\bibitem[{{Heesen} {et~al.}(2011){Heesen}, {Rau}, {Rupen}, {Brinks}, \&
  {Hunter}}]{Heesen11}
{Heesen}, V., {Rau}, U., {Rupen}, M.~P., {Brinks}, E., \& {Hunter}, D.~A. 2011,
  \apjl, 739, L23

\bibitem[{{Helou} {et~al.}(1985){Helou}, {Soifer}, \&
  {Rowan-Robinson}}]{Helou85}
{Helou}, G., {Soifer}, B.~T., \& {Rowan-Robinson}, M. 1985, \apjl, 298, L7

\bibitem[{{Ivison} {et~al.}(2010){Ivison}, {Magnelli}, {Ibar}, {Andreani},
  {Elbaz}, {Altieri}, {Amblard}, {Arumugam}, {Auld}, {Aussel}, {Babbedge},
  {Berta}, {Blain}, {Bock}, {Bongiovanni}, {Boselli}, {Buat}, {Burgarella},
  {Castro-Rodr{\'{\i}}guez}, {Cava}, {Cepa}, {Chanial}, {Cimatti}, {Cirasuolo},
  {Clements}, {Conley}, {Conversi}, {Cooray}, {Daddi}, {Dominguez}, {Dowell},
  {Dwek}, {Eales}, {Farrah}, {F{\"o}rster Schreiber}, {Fox}, {Franceschini},
  {Gear}, {Genzel}, {Glenn}, {Griffin}, {Gruppioni}, {Halpern},
  {Hatziminaoglou}, {Isaak}, {Lagache}, {Levenson}, {Lu}, {Lutz}, {Madden},
  {Maffei}, {Magdis}, {Mainetti}, {Maiolino}, {Marchetti}, {Morrison},
  {Mortier}, {Nguyen}, {Nordon}, {O'Halloran}, {Oliver}, {Omont}, {Owen},
  {Page}, {Panuzzo}, {Papageorgiou}, {Pearson}, {P{\'e}rez-Fournon}, {P{\'e}rez
  Garc{\'{\i}}a}, {Poglitsch}, {Pohlen}, {Popesso}, {Pozzi}, {Rawlings},
  {Raymond}, {Rigopoulou}, {Riguccini}, {Rizzo}, {Rodighiero}, {Roseboom},
  {Rowan-Robinson}, {Saintonge}, {Sanchez Portal}, {Santini}, {Schulz},
  {Scott}, {Seymour}, {Shao}, {Shupe}, {Smith}, {Stevens}, {Sturm},
  {Symeonidis}, {Tacconi}, {Trichas}, {Tugwell}, {Vaccari}, {Valtchanov},
  {Vieira}, {Vigroux}, {Wang}, {Ward}, {Wright}, {Xu}, \& {Zemcov}}]{Ivison10a}
{Ivison}, R.~J., {Magnelli}, B., {Ibar}, E., {et~al.} 2010, \aap, 518, L31

\bibitem[{{Jarvis} {et~al.}(2010){Jarvis}, {Smith}, {Bonfield}, {Hardcastle},
  {Falder}, {Stevens}, {Ivison}, {Auld}, {Baes}, {Baldry}, {Bamford}, {Bourne},
  {Buttiglione}, {Cava}, {Cooray}, {Dariush}, {de Zotti}, {Dunlop}, {Dunne},
  {Dye}, {Eales}, {Fritz}, {Hill}, {Hopwood}, {Hughes}, {Ibar}, {Jones},
  {Kelvin}, {Lawrence}, {Leeuw}, {Loveday}, {Maddox}, {Micha{\l}owski},
  {Negrello}, {Norberg}, {Pohlen}, {Prescott}, {Rigby}, {Robotham},
  {Rodighiero}, {Scott}, {Sharp}, {Temi}, {Thompson}, {van der Werf}, {van
  Kampen}, {Vlahakis}, \& {White}}]{Jarvis10}
{Jarvis}, M.~J., {Smith}, D.~J.~B., {Bonfield}, D.~G., {et~al.} 2010, \mnras,
  409, 92

\bibitem[{{Kazantsev}(1968)}]{Kazantsev68}
{Kazantsev}, A.~P. 1968, Sov. Phys. JETP, 26, 1031

\bibitem[{{Kennicutt} \& {Evans}(2012)}]{Kennicutt12}
{Kennicutt}, R.~C. \& {Evans}, N.~J. 2012, \araa, 50, 531

\bibitem[{{Kennicutt}(1998)}]{Kennicutt98}
{Kennicutt}, Jr., R.~C. 1998, \apj, 498, 541

\bibitem[{{Kennicutt} {et~al.}(2008){Kennicutt}, {Lee}, {Funes}, {Sakai}, \&
  {Akiyama}}]{Kennicutt08}
{Kennicutt}, Jr., R.~C., {Lee}, J.~C., {Funes}, Jos{\'e}~G., S.~J., {Sakai},
  S., \& {Akiyama}, S. 2008, \apjs, 178, 247

\bibitem[{{Kepley} {et~al.}(2010){Kepley}, {M{\"u}hle}, {Everett}, {Zweibel},
  {Wilcots}, \& {Klein}}]{Kepley10}
{Kepley}, A.~A., {M{\"u}hle}, S., {Everett}, J., {et~al.} 2010, ApJ, 712, 536

\bibitem[{{Kepley} {et~al.}(2011){Kepley}, {Zweibel}, {Wilcots}, {Johnson}, \&
  {Robishaw}}]{Kepley11}
{Kepley}, A.~A., {Zweibel}, E.~G., {Wilcots}, E.~M., {Johnson}, K.~E., \&
  {Robishaw}, T. 2011, ApJ, 736, 139

\bibitem[{{Kroupa}(2002)}]{Kroupa02}
{Kroupa}, P. 2002, Science, 295, 82

\bibitem[{{Kr{\"u}gel}(2008)}]{Krugel08}
{Kr{\"u}gel}, E. 2008, {An introduction to the physics of interstellar dust.
  Taylor Francis, Series in astronomy and astrophysics.}

\bibitem[{{Lacki} \& {Thompson}(2010)}]{Lacki10b}
{Lacki}, B.~C. \& {Thompson}, T.~A. 2010, \apj, 717, 196

\bibitem[{{Lacki} {et~al.}(2010){Lacki}, {Thompson}, \& {Quataert}}]{Lacki10}
{Lacki}, B.~C., {Thompson}, T.~A., \& {Quataert}, E. 2010, \apj, 717, 1

\bibitem[{{Latif} {et~al.}(2013){Latif}, {Schleicher}, {Schmidt}, \&
  {Niemeyer}}]{Latif13}
{Latif}, M.~A., {Schleicher}, D.~R.~G., {Schmidt}, W., \& {Niemeyer}, J. 2013,
  \mnras, 432, 668

\bibitem[{{Lee} {et~al.}(2009){Lee}, {Gil de Paz}, {Tremonti}, {Kennicutt},
  {Salim}, {Bothwell}, {Calzetti}, {Dalcanton}, {Dale}, {Engelbracht}, {Funes},
  {Johnson}, {Sakai}, {Skillman}, {van Zee}, {Walter}, \& {Weisz}}]{Lee09}
{Lee}, J.~C., {Gil de Paz}, A., {Tremonti}, C., {et~al.} 2009, \apj, 706, 599

\bibitem[{{Lisenfeld} \& {V{\"o}lk}(2010)}]{Lisenfeld10}
{Lisenfeld}, U. \& {V{\"o}lk}, H.~J. 2010, \aap, 524, A27

\bibitem[{{Mao} {et~al.}(2008){Mao}, {Gaensler}, {Stanimirovi{\'c}},
  {Haverkorn}, {McClure-Griffiths}, {Staveley-Smith}, \& {Dickey}}]{Mao08}
{Mao}, S.~A., {Gaensler}, B.~M., {Stanimirovi{\'c}}, S., {et~al.} 2008, \apj,
  688, 1029

\bibitem[{{McKee} \& {Ostriker}(1977)}]{McKee77}
{McKee}, C.~F. \& {Ostriker}, J.~P. 1977, \apj, 218, 148

\bibitem[{{McKenzie} \& {Webb}(1984)}]{McKenzie84}
{McKenzie}, J.~F. \& {Webb}, G.~M. 1984, Journal of Plasma Physics, 31, 275

\bibitem[{{Miettinen} {et~al.}(2015){Miettinen}, {Novak}, {Smol{\v c}i{\'c}},
  {Schinnerer}, {Sargent}, {Murphy}, {Aravena}, {Bondi}, {Carilli}, {Karim},
  {Salvato}, \& {Zamorani}}]{Miettinen15}
{Miettinen}, O., {Novak}, M., {Smol{\v c}i{\'c}}, V., {et~al.} 2015, \aap, 584,
  A32

\bibitem[{{Mulcahy} {et~al.}(2014){Mulcahy}, {Horneffer}, {Beck}, {Heald},
  {Fletcher}, {Scaife}, {Adebahr}, {Anderson}, {Bonafede}, {Br{\"u}ggen},
  {Brunetti}, {Chy{\.z}y}, {Conway}, {Dettmar}, {En{\ss}lin}, {Haverkorn},
  {Horellou}, {Iacobelli}, {Israel}, {Junklewitz}, {Jurusik}, {K{\"o}hler},
  {Kuniyoshi}, {Orr{\'u}}, {Paladino}, {Pizzo}, {Reich}, \&
  {R{\"o}ttgering}}]{Mulcahy14}
{Mulcahy}, D.~D., {Horneffer}, A., {Beck}, R., {et~al.} 2014, \aap, 568, A74

\bibitem[{{Murphy}(2009)}]{Murphy09}
{Murphy}, E.~J. 2009, \apj, 706, 482

\bibitem[{{Murphy}(2013)}]{Murphy13}
{Murphy}, E.~J. 2013, \apj, 777, 58

\bibitem[{{Murphy} {et~al.}(2006){Murphy}, {Helou}, {Braun}, {Kenney}, {Armus},
  {Calzetti}, {Draine}, {Kennicutt}, {Roussel}, {Walter}, {Bendo}, {Buckalew},
  {Dale}, {Engelbracht}, {Smith}, \& {Thornley}}]{Murphy06}
{Murphy}, E.~J., {Helou}, G., {Braun}, R., {et~al.} 2006, \apjl, 651, L111

\bibitem[{{Niklas}(1997)}]{Niklas97}
{Niklas}, S. 1997, \aap, 322, 29

\bibitem[{{Niklas} \& {Beck}(1997)}]{Niklas97b}
{Niklas}, S. \& {Beck}, R. 1997, A\&A, 320, 54

\bibitem[{{Oman} {et~al.}(2015){Oman}, {Navarro}, {Fattahi}, {Frenk}, {Sawala},
  {White}, {Bower}, {Crain}, {Furlong}, {Schaller}, {Schaye}, \&
  {Theuns}}]{Oman15}
{Oman}, K.~A., {Navarro}, J.~F., {Fattahi}, A., {et~al.} 2015, \mnras, 452,
  3650

\bibitem[{{Roychowdhury} \& {Chengalur}(2012)}]{Roychowdhury12}
{Roychowdhury}, S. \& {Chengalur}, J.~N. 2012, \mnras, 423, L127

\bibitem[{{Roychowdhury} {et~al.}(2009){Roychowdhury}, {Chengalur}, {Begum}, \&
  {Karachentsev}}]{Roychowdhury09}
{Roychowdhury}, S., {Chengalur}, J.~N., {Begum}, A., \& {Karachentsev}, I.~D.
  2009, \mnras, 397, 1435

\bibitem[{{Roychowdhury} {et~al.}(2015){Roychowdhury}, {Huang}, {Kauffmann},
  {Wang}, \& {Chengalur}}]{Roychowdhury15}
{Roychowdhury}, S., {Huang}, M.-L., {Kauffmann}, G., {Wang}, J., \&
  {Chengalur}, J.~N. 2015, \mnras, 449, 3700

\bibitem[{{Salpeter}(1955)}]{Salpeter55}
{Salpeter}, E.~E. 1955, \apj, 121, 161

\bibitem[{{Sargent} {et~al.}(2010){Sargent}, {Schinnerer}, {Murphy}, {Carilli},
  {Helou}, {Aussel}, {Le Floc'h}, {Frayer}, {Ilbert}, {Oesch}, {Salvato},
  {Smol{\v c}i{\'c}}, {Kartaltepe}, \& {Sanders}}]{Sargent10}
{Sargent}, M.~T., {Schinnerer}, E., {Murphy}, E., {et~al.} 2010, \apjl, 714,
  L190

\bibitem[{{Schekochihin} {et~al.}(2002){Schekochihin}, {Cowley}, {Hammett},
  {Maron}, \& {McWilliams}}]{Scheko02}
{Schekochihin}, A.~A., {Cowley}, S.~C., {Hammett}, G.~W., {Maron}, J.~L., \&
  {McWilliams}, J.~C. 2002, New Journal of Physics, 4, 84

\bibitem[{{Schlei\-cher} {et~al.}(2010){Schlei\-cher}, {Banerjee}, {Sur},
  {Arshakian}, {Klessen}, {Beck}, \& {Spaans}}]{Schleicher10c}
{Schlei\-cher}, D.~R.~G., {Banerjee}, R., {Sur}, S., {et~al.} 2010, A\&A, 522,
  A115

\bibitem[{{Schleicher} \& {Beck}(2013)}]{Schleicher13b}
{Schleicher}, D.~R.~G. \& {Beck}, R. 2013, \aap, 556, A142

\bibitem[{{Schleicher} {et~al.}(2013){Schleicher}, {Schober}, {Federrath},
  {Bovino}, \& {Schmidt}}]{Schleicher13}
{Schleicher}, D.~R.~G., {Schober}, J., {Federrath}, C., {Bovino}, S., \&
  {Schmidt}, W. 2013, New Journal of Physics, 15, 023017

\bibitem[{{Schmidt}(1959)}]{Schmidt59}
{Schmidt}, M. 1959, \apj, 129, 243

\bibitem[{{Schober} {et~al.}(2012){Schober}, {Schleicher}, {Federrath},
  {Klessen}, \& {Banerjee}}]{Schober12b}
{Schober}, J., {Schleicher}, D., {Federrath}, C., {Klessen}, R., \& {Banerjee},
  R. 2012, PRE, 85, 026303

\bibitem[{{Schober} {et~al.}(2013){Schober}, {Schleicher}, \&
  {Klessen}}]{Schober13}
{Schober}, J., {Schleicher}, D.~R.~G., \& {Klessen}, R.~S. 2013, \aap, 560, A87

\bibitem[{{Schober} {et~al.}(2015){Schober}, {Schleicher}, \&
  {Klessen}}]{Schober15}
{Schober}, J., {Schleicher}, D.~R.~G., \& {Klessen}, R.~S. 2015, \mnras, 446, 2

\bibitem[{{Shu} {et~al.}(2005){Shu}, {Mo}, \& {Shu-DeMao}}]{Shu05}
{Shu}, C.-G., {Mo}, H.-J., \& {Shu-DeMao}. 2005, \cjaa, 5, 327

\bibitem[{{Subramanian}(1999)}]{Subramanian99}
{Subramanian}, K. 1999, Physical Review Letters, 83, 2957

\bibitem[{{Swaters} {et~al.}(2009){Swaters}, {Sancisi}, {van Albada}, \& {van
  der Hulst}}]{Swaters09}
{Swaters}, R.~A., {Sancisi}, R., {van Albada}, T.~S., \& {van der Hulst}, J.~M.
  2009, \aap, 493, 871

\bibitem[{{Tabatabaei} {et~al.}(2013){Tabatabaei}, {Schinnerer}, {Murphy},
  {Beck}, {Groves}, {Meidt}, {Krause}, {Rix}, {Sandstrom}, {Crocker},
  {Galametz}, {Helou}, {Wilson}, {Kennicutt}, {Calzetti}, {Draine}, {Aniano},
  {Dale}, {Dumas}, {Engelbracht}, {Gordon}, {Hinz}, {Kreckel}, {Montiel}, \&
  {Roussel}}]{Taba13}
{Tabatabaei}, F.~S., {Schinnerer}, E., {Murphy}, E.~J., {et~al.} 2013, \aap,
  552, A19

\bibitem[{{Theis} \& {Kohle}(2001)}]{Theis01}
{Theis}, C. \& {Kohle}, S. 2001, \aap, 370, 365

\bibitem[{{Tielens}(2005)}]{Tielens05}
{Tielens}, A.~G.~G.~M. 2005, {The Physics and Chemistry of the Interstellar
  Medium}

\bibitem[{{van der Kruit}(1973{\natexlab{a}})}]{Kruit73b}
{van der Kruit}, P.~C. 1973{\natexlab{a}}, A\&A, 29, 249

\bibitem[{{van der Kruit}(1973{\natexlab{b}})}]{Kruit73c}
{van der Kruit}, P.~C. 1973{\natexlab{b}}, A\&A, 29, 263

\bibitem[{{van der Kruit}(1973{\natexlab{c}})}]{Kruit73a}
{van der Kruit}, P.~C. 1973{\natexlab{c}}, A\&A, 29, 231

\bibitem[{{V\"olk}(1989)}]{Volk89}
{V\"olk}, H.~J. 1989, A\&A, 218, 67

\bibitem[{{Walter} {et~al.}(2008){Walter}, {Brinks}, {de Blok}, {Bigiel},
  {Kennicutt}, {Thornley}, \& {Leroy}}]{Walter08}
{Walter}, F., {Brinks}, E., {de Blok}, W.~J.~G., {et~al.} 2008, AJ, 136, 2563

\bibitem[{{Wang} \& {Abel}(2009)}]{Wang09}
{Wang}, P. \& {Abel}, T. 2009, ApJ, 696, 96

\bibitem[{{Weidner} \& {Kroupa}(2005)}]{Weidner05}
{Weidner}, C. \& {Kroupa}, P. 2005, \apj, 625, 754

\bibitem[{{Yun} {et~al.}(2001){Yun}, {Reddy}, \& {Condon}}]{Yun01}
{Yun}, M.~S., {Reddy}, N.~A., \& {Condon}, J.~J. 2001, \apj, 554, 803

\end{thebibliography}

\end{document}